%% file: ms.tex
\documentclass[preprint]{aastex}
\usepackage{amsmath,amsfonts}
\input mathdef.tex

\begin{document}
\title{Frequency decomposition of astrometric signature of planetary
systems}
\author{Maciej Konacki \altaffilmark{1}, 
\affil{Department of Geological and Planetary Sciences, California
Institute of Technology, 1201 E. California Blvd., Pasadena, CA 91125, USA}
Andrzej J.~Maciejewski  \altaffilmark{2}}
\affil{Toru\'n Centre for Astronomy, 
           Nicolaus Copernicus University,\\   
           87-100 Toru\'n, Gagarina 11, Poland}
\author{Alex Wolszczan \altaffilmark{3}}
\affil{Department of Astronomy and Astrophysics, Penn State University \\
University Park, PA 16802, USA \\
Toru\'n Centre for Astronomy, Nicolaus Copernicus University \\
ul. Gagarina 11, 87-100 Toru\' n, Poland}
\altaffiltext{1}{e-mail: maciej@gps.caltech.edu}
\altaffiltext{2}{e-mail: maciejka@astri.uni.torun.pl}
\altaffiltext{3}{e-mail: alex@astro.psu.edu}
\begin{abstract}
We present theoretical analysis of the astrometric searches for
extrasolar planets with the Space Interferometry Mission (SIM). 
Particularly, we derive a model for the future measurements with SIM 
and discuss the problem of reliable estimation of orbital elements
of planets. For this purpose we propose a new method of data analysis 
and present a numerical test of its application on simulated
SIM astrometric measurements of $\upsilon$ Andromedae planetary system.
We demonstrate that our approach allows successfull determination
of its orbital elements.
\end{abstract}

\section{Introduction}

One of the most important and challenging goals of the Space Interferometry 
Mission (SIM, see {\tt http://sim.jpl.nasa.gov/}) is astrometric detection 
of extrasolar planetary systems including Earth-like planets around stars 
from the solar neighborhood. High precision astrometry requires not only 
advanced technology but also adequately elaborated methods of data analysis. 
Among others it is important to develop
techniques allowing reliable detection of planetary signatures and
extraction of the orbital elements. The aim of this paper is to discuss
some of the problems related to this subject. Specifically, we propose 
a method called the Frequency Decomposition (FD) to detect planets
and help to obtain their orbital elements. This method has been 
successfully used for PSR 1257+12 timing observations \citep{Konacki:99a::} 
and 16 Cygni B radial velocity measurements \citep{Konacki:99b::}. Particular 
nature of the astrometric observations requires however some modifications of 
our original approach. In the paper we present a theoretical background of 
the method and an example of its application. 

The astrometric signal is a superposition of several effects of different 
magnitude and a proper analysis of the observations requires at least rough 
a priori knowledge of how different effects contribute to the signal. 
However, these effects depend on the parameters (such as number of planets,
their eccentricities) which are unknown in advance. So what we propose is a
two-step analysis (1) FD to understand the basic properties of the signal
(i.e determine the number of planets and approximate values of their orbital 
elements) and (2) least-squares fit based on a proper model and the starting 
values of parameters derived from the previous step to refine the parameters 
and obtain their uncertainties.

The basic idea of FD is the following. With few exception (proper motion, 
long period planets) the processes contributing to the signal are periodic.
Therefore our astrometric signal can be successfully modeled as a multiple
Fourier series plus a polynomial of certain degree (to account for the 
proper motion and long period planets). FD is a numerical algorithm to
obtain the estimates of frequencies, amplitudes and phases of such model 
\citep{Konacki:99a::}. Let us note that such approach, contrary to the usual
least-squares method, allows us to analyze the data without assuming any 
physical model. Subsequently we interpret derived parameters. We decide how 
many planets are present in the system and calculate their orbital
parameters (as one can derive analytical formula expressing amplitudes and 
phases as functions of the orbital elements). This is especially useful for 
multiple planetary systems where deciphering the number of planets may be 
tricky \citep[e.g. two planets in circular 2:1 resonant orbits may mimic one 
planet in an eccentric orbit, see][]{Konacki:99c::}. Our approach can be also 
helpful while trying to determine whether we observe an astrometric
displacement from a planet in 1-yr orbit or annual parallax since the
parallactic motion has its own specific Fourier expansion constrained 
by SIM orbit. Finally, we can use these findings to perform the 'traditional' 
least-squares fit. We believe that such approach allows us to make more 
justified hypothesis about the data and in consequence lead to reliable 
results. 

The plan of our paper is the following. In section 2 we derive a detailed
model of the SIM measurements. In section 3 we investigate Fourier
properties of the orbital motion. In section 4 we discuss our approach
to the analysis of SIM data and finally in section 5 we perform some 
numerical tests to show how our method works in practice.

\section{Modeling delays}

SIM measures relative positions of stars using Michelson interferometers.
A single measurement with SIM gives the projection of direction to the star 
$\bs$ onto the interferometer baseline vector $\bB$. The measured quantity is 
the optical pathlength delay between the two arms of the interferometer
\citep{Shao:99::}
\begin{equation}
  \label{eq:delay}
  d = \bB\cdot\bs + c + \epsilon, 
\end{equation}
where $c$ is the zero point of the metrology gauge and $\epsilon$
represents measurement uncertainty. The search for
extrasolar planet is performed in so-called narrow angle mode where
delays toward two stars (called target and reference) within $1^\circ$ 
are measured and compared. For this kind of observation the measured 
quantity becomes the relative delay
\begin{equation}
  \label{eq:reldelay}
   D = \bB\cdot(\bs_1-\bs_2) + \epsilon, 
\end{equation}
where $\bs_1$ and $\bs_2$ are directions to the target and reference
stars, respectively. Such narrow-angle measurement gives the angular 
separation between the stars and offers higher accuracy as many 
errors scale with the angular distance.

The direction to a star $\bS=\bS(t)$ from the Solar System Barycenter
(SSB) is changing with time due to the proper motion and presence of 
companions. These two effects, we model in the following way
\begin{equation}
  \label{eq:S(t)}
  \bS(t) = \bS_0 + \delta\bS_\mu(t) + \delta\bS_{\mathrm c}(t),
\end{equation}
where $\bS_0$ is the direction toward the star at epoch $t_0$;
$\delta\bS_\mu(t)$ and $ \delta\bS_{\mathrm c}(t)$ describe changes
due to the proper and orbital motion, respectively.  In order to properly 
calculate these changes let us assume that the SSB radius vector
of each star is given by
\begin{equation}
  \label{eq:rr}
   \bR = \bR_0 + \delta\bR, \qquad \text{where}\quad 
    \norm{\delta\bR}\ll\norm{\bR_0}, 
\end{equation}
then up to the first order in $\norm{\delta\bR}/\norm{\bR_0}$
\begin{equation}
  \label{eq:ss}
   \bS = \frac{\bR}{\norm{\bR}} = \bS_0 + \delta \bS^{(1)}, 
\end{equation}
where
\begin{equation}
  \label{eq:dum}
   \bS_0 = \frac{\bR_0}{\norm{\bR_0}}, \qquad 
   \delta\bS^{(1)} = -\frac{1}{\norm{\bR_0}} \bS_0
   \times\left( \bS_0\times \delta\bR\right), 
\end{equation}
The correction $\delta\bS^{(1)}$ can be written in the form
\begin{equation}
  \label{eq:ds}
  \delta\bS^{(1)} = \frac{1}{\norm{\bR_0}}\left[ \delta\bR - \bS_0
  (\bS_0\cdot\delta\bR)\right] =  \frac{1}{\norm{\bR_0}}\delta\bR_\perp,
\end{equation}
which means that it only depends on the component of $\delta\bR$ 
perpendicular to $\bS_0$. It turns out that sometimes the first
order correction is not sufficient. Therefore we need to derive
and analyze also the second order term $\delta\bS^{(2)}$ given by
\begin{equation}
  \label{eq:ds2} 
  \delta\bS^{(2)} = \bS_0\left[ \frac{3}{2}\left( \bS_0 \cdot
\frac{\delta\bR}{\norm{\bR_0}}\right)^2 -
\frac{1}{2}\left(\frac{\norm{\delta\bR}}{\norm{\bR_0}}\right)^2\right]
- \frac{\delta\bR}{\norm{\bR_0}}\left( \bS_0 \cdot
\frac{\delta\bR}{\norm{\bR_0}}\right)
\end{equation}
If we represent $\delta\bR$ as a sum of two components, perpendicular
and parallel to $\bS_0$,
\begin{equation} 
\delta\bR = \delta\bR_\perp + \delta\bR_\parallel
\end{equation}
$\delta\bS^{(2)}$ can be written as
\begin{equation}
  \label{eq:ds2:1}
\delta\bS^{(2)} = -\frac{1}{2}
\left(\frac{\norm{\delta\bR_\perp}}{\norm{\bR_0}}\right)^2\bS_0
-\frac{\norm{\delta\bR_\parallel}}{\norm{\bR_0}^2}\delta\bR_\perp
\end{equation}
As we show in section 2.3, $\delta\bS^{(2)}$ is especially 
significant for nearby stars with large proper motions. For such
stars we have
\begin{equation}
  \label{vel}
\delta\bR_\perp = \bV_T\,t \quad \delta\bR_\parallel = \bV_R\,t
\end{equation}                                                   
where $\bV_T$ and $\bV_R$ are respectively transverse and radial 
velocity of the star. Thus if the star has a significant proper motion, 
through astrometric observations we can detect angular displacement 
$\delta \theta$ (second term in equation \eqref{eq:ds2:1})
\begin{equation}
\label{eq:ang2}
\delta \theta = \frac{\norm{\delta\bR_\parallel}}{\norm{\bR_0}}
\frac{\norm{\delta\bR_\perp}}{\norm{\bR_0}} = 
\frac{\norm{\bV_T}}{\norm{\bR_0}}
\frac{\norm{\bV_R}}{\norm{\bR_0}}\,t^2
\end{equation} 
due to the radial velocity. This effect is called {\it perspective 
acceleration}. The other term from the equation \eqref{eq:ds2:1}
has an interesting property. Namely, it can be shown that it does not
change the angle between $\bS(t) = \bS_0 + \delta\bS^{(1)}(t) +
\delta\bS^{(2)}(t)$ and $\bS_0$ (i.e. current and initial position of 
the star). In other words, if we had a direct way to measure this 
angle, we would not observe any effect from that term. However,
since we measure all angles through the equation \eqref{eq:delay}
and model the unit vector toward the star with 
$\bS(t) = \bS_0 + \delta\bS^{(1)}(t) + \delta\bS^{(2)}(t)$, the term
$-\frac{1}{2}\left(\norm{\delta\bR_\perp}/\norm{\bR_0}\right)^2\bS_0$
is necessary. Specifically it affects the length of the vector
$\bS$ and helps to keep it normalized within the accuracy of the second
order approximation. Further details concerning second order
corrections we discuss in section 2.3.

\subsection{Local frame and baseline vector orientations}

In order to obtain explicit form of the delay $d$ we need to
calculate a scalar product $\bB\cdot\bS$. The value of this product does
not depend on a chosen reference frame. Thus, depending on our needs
we can express vectors on the right hand side of \eqref{eq:S(t)} in 
different ways. It is convenient to introduce a local right hand 
orthonormal frame at the point $\bS_0$ on the celestial sphere (see Fig.~1).  
This frame is connected with the classical equatorial spherical 
coordinates and is defined by the unit vectors $\{\be_\alpha, \be_\delta, 
\be_r\}$. In SSB equatorial frame coordinates of these vectors are 
the following
\begin{gather}
  \label{eq:eeS}
  \be_\alpha=(-\sin\alpha,\cos\alpha,0), \qquad
  \be_\delta=(-\sin\delta\cos\alpha,-\sin\delta\sin\alpha,\cos\delta),\\
  \be_r= \bS_0=(\cos\delta\cos\alpha,\cos\delta\sin\alpha,\sin\delta),
\end{gather}
where $(\alpha,\delta)= (\alpha_0,\delta_0)$ is the right ascension
and declination of a star at $t_0$. 

One can determine the relative position of the target and reference star 
using two interferometers or, as it is planned for SIM, by performing two 
measurements with one interferometer for two non parallel orientations of 
its baseline, $\bB_i,\,i=1,2$. For each orientation, the baseline can be 
represented as a sum of two vectors, $\bB_i^{\shortparallel}, \bB_i^{\perp}$, 
parallel and perpendicular to the initial direction toward the target star, 
$\bS_0$. Since we have $\bS(t) = \bS_0 + \delta\bS(t)$ where $\delta\bS(t)$ 
is a displacement tangent to $\bS_0$, the delay can be written as
\begin{equation}  
\label{eq:base12}
d = \bB_{i}\cdot\bS(t) + c + \epsilon = \bB_i^{\shortparallel}\cdot\bS_0 + 
\bB_i^{\perp}\cdot\delta\bS(t) + c + \epsilon = d_0 + \Delta d(t) + c +
\epsilon
\end{equation}
where $d_0 = \bB_i^{\shortparallel}\cdot\bS_0$ and $\Delta d(t) =
\bB_i^{\perp}\cdot\delta\bS(t)$. Clearly, from the planet detection point of 
view the important term is $\Delta d(t)$. Assuming that the measurement 
uncertainty, $\epsilon$, is independent on the baseline orientation the most 
favorable orientation of $\bB_i$ is $\bB_i = \bB_i^{\perp}$. For such 
orientation $\Delta d(t)$ is the largest possible for a given length of the 
baseline. Moreover, the baseline orientations should be perpendicular. This
way the the covariance ellipse on the sky will be circular and there will not
be any direction on the plane tangent at $\bS_0$ in which the measurements
are more accurate than in others. Therefore, for all further 
considerations we assume that all observations are made with two orthogonal 
and fixed baseline orientations $\bB_1$ and $\bB_2$ which are perpendicular 
to the initial direction toward the target star. Additionally, to simplify
the equations we assume that $\bB_1$ is parallel to $\be_\alpha$ and
$\bB_2$ is parallel to $\be_\delta$.

\subsection{Proper motion, parallax and companions}

The proper motion is a projection on the sky of the motion of a star with 
the velocity $\bV$ and within the first order approximation astrometrically 
only its transverse component $\bV_T = V_{\alpha}\,\be_\alpha + V_{\delta}\,\be_\delta$ 
is observable. Thus using simple arguments we find that
\begin{equation}
\label{eq:dSm}
\delta\bS_\mu(t) = \pi\left(V_{\alpha}\,t\,\be_\alpha + 
V_{\delta}\,t\,\be_\delta\right) =
\cos\delta\, \mu_\alpha\,t\,  \be_\alpha + \mu_\delta\,t \,\be_\delta,   
\end{equation}
where
\begin{equation}
V_{\alpha} = \bV\cdot\be_\alpha, \quad V_{\delta} = \bV\cdot\be_\delta
\quad \text{and} \quad
\mu_\alpha = \frac{\rmd \alpha}{\rmd t}(t_0),
\quad 
\mu_\delta = \frac{\rmd \delta}{\rmd t}(t_0).
\end{equation}
and $\pi= 1/D^\star$ where $D^\star = \norm{\bR_0}$ is the SSB distance to 
the star. 

If the star has companions then the proper motion refers to the
motion of the mass center of the system and the first order correction
due to the orbital is given by the following equation
\begin{equation}
  \label{eq:comp}
  \delta\bS_{\mathrm{c}}(t) = \pi\left[ R^*_\alpha(t) \be_\alpha +
 R^*_\delta(t) \be_\delta\right],
\end{equation}
where $\bR^\star = (R_\alpha^\star,R_\delta^\star,R^\star_r)$ denotes 
the radius vector of the star with respect to the barycenter of its
system in the local frame $\{\be_\alpha, \be_\delta, \be_r\}$. 

If the interferometer is located at $\bR_{\mathrm{O}}(t)$ in 
SSB frame then the observed direction toward the star is
\begin{equation}
 \label{eq:sobs}
 \bs(t) = \bS(t) + \bPi(t),
\end{equation}
where $\bPi(t)$ is the parallactic displacement
\begin{equation}
  \label{eq:parallax}
  \bPi(t) = \pi\bS_0\times\left(\bS_0\times \bR_{\mathrm{O}}(t)\right)
\end{equation}
obtained from the equation \eqref{eq:dum} by substituting $-\bR_{\mathrm{O}}$
for $\delta\bR$. In the local frame the parallactic displacement can 
be written in the following form
\begin{equation}
  \label{eq:plxl}
   \bPi(t) = \pi\left[ \Pi_\alpha(t)\be_\alpha +
    \Pi_\delta(t) \be_\delta \right],
\end{equation}
where
\begin{gather} 
  \Pi_{\alpha}(t) = -\bR_{\mathrm{O}}(t)\cdot\be_\alpha =
X_{\mathrm{O}}(t)\sin\alpha -
  Y_{\mathrm{O}}(t)\cos\alpha, \\
  \Pi_{\delta}(t) = -\bR_{\mathrm{O}}(t)\cdot\be_\delta =
X_{\mathrm{O}}(t)\sin\delta\cos\alpha +
  Y_{\mathrm{O}}(t)\sin\delta\sin\alpha - Z_{\mathrm{O}}(t)\cos\delta
\end{gather}
and $(X_{\mathrm{O}}(t),Y_{\mathrm{O}}(t),Z_{\mathrm{O}}(t))$ are the 
coordinates of SSB vector $\bR_{\mathrm{O}}(t)$. The expressions for
$\bS_{\mu}(t)$, $\bS_{\mathrm{c}}(t)$ and $\bPi(t)$ come directly from 
the formulae \eqref{eq:dum}, \eqref{eq:ds} and thus represent a first
order approximation with respect to $\pi$.

Now, using \eqref{eq:S(t)}--\eqref{eq:parallax} and assuming 
that the measurements have been already corrected for 
aberration and gravitational lensing we can rewrite the delay
equation \eqref{eq:delay} in the form
\begin{equation}
  \label{eq:d}
  d = d^0 + d^\mu\,t + \bd^\pi\cdot\bR_{\mathrm{O}}(t) + 
  \bd^\mathrm{c}\cdot\bR^*(t) + c + \epsilon, 
\end{equation}
where
\begin{gather}
\label{eq:dd}
d^0 = \bB \cdot \bS_0, \qquad d^\mu = \pi\bB\cdot\bV_T = \bb\cdot \bmu, \qquad 
\bmu = ( \mu_\alpha \cos\delta, \mu_\delta), \\
\bb = (\bB\cdot\be_\alpha,\bB\cdot\be_\delta),
\qquad
\bd^\pi = -\pi\left[\bB - (\bB\cdot\bS_0)\bS_0 \right],\\
\bd^\mathrm{c}= \pi\bb, 
\qquad \bR^\star(t)= ( R^\star_\alpha (t) , R^\star_\delta (t)).
\end{gather}
Using \eqref{eq:d} we obtain similar formula for the relative delay
\begin{equation}
  \label{eq:reld}
  D = D^0 + D^\mu\,t + \bD^\pi\cdot\bR_{\mathrm{O}}(t) + 
  \bd^{\mathrm{c}}\cdot \bR^\star(t) + \epsilon,
\end{equation}
where
\begin{equation}
\label{eq:D}
\begin{split}
  D^0 = \bB \cdot( \bS_0^1 -\bS_0^2), \qquad
  D^\mu = \bB\cdot\left(\pi_1\bV_T^1 - \pi_2\bV_T^2\right),\\
  \bD^\pi = 
 \left[\pi_1 (\bB\cdot\bS_0^1)\bS_0^1-\pi_2 (\bB\cdot\bS_0^2)\bS_0^2\right] 
- (\pi_1-\pi_2)\bB.
\end{split}
\end{equation}
where the indices $1,2$ refer to the target and reference star
respectively. In the above we assumed that only the target star has 
companions so as $\bd^{\mathrm{c}}$ refers to the local reference frame 
of the target star. Formula \eqref{eq:reld} plays the 
fundamental role in our consideration. It explicitly shows the structure of 
the observed signal that consists of a dominant linear trend modulated by
periodicities due to the motion of the interferometer and the companions of 
the target star.

According to our assumptions a single observation is done for two
orthogonal baseline orientations $\bB_1$ and $\bB_2$. Thus a single
observations is given as a two component vector $\bD=(D_1,D_2)$ of the
relative delays. Each component $D_i$ has the form \eqref{eq:reld} where 
coefficients $D^0_i$, $D^\mu_i$, $\bD^{\pi}_i$ and $\bd^{\mathrm{c}}_i$ 
are calculated with the formulae \eqref{eq:D} and $\bB=\bB_i$, $i=1,2$ 
respectively.

\subsection{Second order corrections}

The above considerations represent first order approximation which is 
sufficient for most astrometric measurements. However SIM is expected to 
deliver unprecedented 1$\mu$as precision in the narrow-angle mode and thus 
it is important to understand limitation of the model \eqref{eq:D}. It can be 
accomplished by analyzing higher order terms. For the baselines
perpendicular to $\bS_0$, the second order corrections correspond
to the term that is responsible for the actual angular displacement
(see equation \eqref{eq:ang2}) and we have 
\begin{equation}  
\label{eq:corr3}
\norm{\delta\bS^{(2)}} = \frac{ \norm{\delta\bR_{\shortparallel}} }
{ \norm{\bR_0}}\frac{ \norm{\delta\bR_{\perp}} }{ \norm{\bR_0}
} \le \left(\frac{ \norm{\delta\bR} }{ \norm{\bR_0} }\right)^2 
\end{equation}
They can be calculated if we put $\delta\bR = -\bR_{\mathrm{O}}(t) +
\bR^{\star}(t) + \bR_{V}(t)$ where $\bR_{V}(t) = \bV_T\,t + \bV_R\,t$.
We obtain
\begin{equation}
\label{eq:corr3a}
\begin{array}{l}
\displaystyle
\frac{1}{\pi^2}\delta\bS^{(2)} = -\norm{\bR_{\mathrm{O}}^{\shortparallel}(t)}
\bR_{\mathrm{O}}^{\perp}(t) + 
\norm{\bR_{\shortparallel}^{\star}(t)}\bR_{\perp}^{\star}(t) +
\norm{\bR_{V}^{\shortparallel}(t)}\bR_{V}^{\perp}(t) + \\[0.5cm]                                   
\displaystyle
+ \, \left(\norm{\bR_{\mathrm{O}}^{\shortparallel}(t)}
\bR_{\perp}^{\star}(t) -
\norm{\bR_{\shortparallel}^{\star}(t)}\bR_{\mathrm{O}}^{\perp}(t) 
\right) + \left(\norm{\bR_{\mathrm{O}}^{\shortparallel}(t)}               
\bR^{\perp}_{V}(t) -
\norm{\bR^{\shortparallel}_{V}(t)}\bR_{\mathrm{O}}^{\perp}(t)
\right) + \\[0.5cm]                                   
\displaystyle
+ \,\left(\norm{\bR_{\star}^{\shortparallel}(t)}               
\bR^{\perp}_{V}(t) +
\norm{\bR^{\shortparallel}_{V}(t)}\bR^{\star}_{\perp}(t)
\right)  
\end{array}
\end{equation}
and
\begin{equation}
\label{eq:corr3b}
\begin{array}{l}
\displaystyle
\displaystyle
\norm{\delta\bS^{(2)}} \le \pi^2\big( \norm{\bR_{\mathrm{O}}(t)}^2
+ \norm{\bR^{\star}(t)}^2 + \norm{\bR_{V}(t)}^2 + \\[0.5cm]
\displaystyle 
-\,2\,\bR_{\mathrm{O}}(t)\cdot\bR^{\star}(t) - 2\,
\bR_{\mathrm{O}}(t)\cdot\bR_{V}(t) + 2\,\bR^{\star}(t)\cdot\bR_{V}(t)
\big)
\end{array}
\end{equation}
As we can see there are two types of second order corrections. The first
type includes the second order corrections due to the proper motion,
parallax and companions. For the proper motion it is easy to calculate
its exact value
\begin{equation}   
\Delta S_{\mu} = \frac{1}{4}\frac{\norm{\bV_T}}{\norm{\bR_0}}
\frac{\norm{\bV_R}}{\norm{\bR_0}}\,\Delta T^2 =
\frac{\pi^2}{4}\,V_T\,V_R\,\Delta T^2
\end{equation}
where $\Delta T$ is the time span of the mission (and we assumed that
$t_0$ is at half of $\Delta T$), $\norm{\bV_T} = V_T$, $\norm{\bV_R} = V_R$ 
In order to learn about the magnitude of this correction, we calculated its 
value for the sample of 150 stars from the Hipparcos catalogue with
the largest proper motion (see the Internet location {\tt 
http://astro.estec.esa.nl/SA-general/Projects/Hipparcos/hipparcos.html}). 
The results are shown in Fig. 3. As one can see $\Delta S_{\mu}$ is indeed 
significant for such stars and without any doubts has to be included into 
the model. For the remaining corrections due to the motion of the 
interferometer (i.e. the second order parallactic correction) and the presence 
of a companion we have the following upper limits
\begin{equation}
  \label{eq:corr4}
\begin{array}{c}
\displaystyle
\pi^2\norm{\bR_{\mathrm{O}}^{\shortparallel}(t)}\norm{\bR_{\mathrm{O}}^{\perp}(t)} \le
\pi^2\norm{\bR_{\mathrm{O}}(t)}^2 \lessapprox \Delta S_{\pi}
= \pi^2 a_{\mathrm{O}}^2, \\[0.4cm]
\displaystyle
\pi^2\norm{\bR_{\shortparallel}^{\star}(t)}\norm{\bR_{\perp}^{\star}(t)} \le
\pi^2\norm{\bR^{\star}(t)}^2 \lessapprox \Delta S_{\mathrm{c}} 
= \pi^2 \left(a_{\star}/(1 - e_{\star})\right)^2
\end{array}
\end{equation}
where $a_{\mathrm{O}}$ is SIM's orbit semi-major axis, $a_{\star}$
semi-major axis of the star's orbit and $e_{\star}$ its eccentricity.
If we assume that SIM semi-major axis is 1 AU then
\begin{equation}  
  \label{eq:corr5}
\begin{array}{c}
\displaystyle
\Delta S_{\mathrm{c}} \approx 5\times10^{-6}\,\frac{m_{M_{JUP}}^2
P_{yr}^{4/3}}{d_{pc}^{2}M_{M_{\odot}}^{4/3}(1 - e)^2}\,\mu as\quad
\mbox{for planetary companions} \\[0.7cm]
\displaystyle
\Delta S_{\mathrm{c}} \approx 4.9 \frac{m_{M_{\odot}}^2
P_{yr}^{4/3}}{d_{pc}^{2} M_{M_{\odot}}^{4/3} (1 + m_{M_{\odot}}
/M_{M_{\odot}})^{4/3}(1 - e)^2}\,\mu as
\quad \mbox{for stellar companions}  \\[0.7cm]
\displaystyle
\Delta S_{\pi} \approx 4.9  \, d_{pc}^{-2} \, \mu as 
\end{array}   
\end{equation}
where $d_{pc}$ is the distance to the star in parsecs, $M_{M_{\odot}}$
mass of the star in solar masses, $m_{M_{\odot}}$ mass of the companion
in solar masses, $m_{M_{JUP}}$ mass of the companion
in Jupiter masses, $P_{yr}$ orbital period in years and $e$ eccentricity. 

The other type of second order corrections includes all "mixed" terms
\begin{equation}
\label{eq:mix}
\begin{array}{c}
\displaystyle
\pi^2\norm{\norm{\bR_{\mathrm{O}}^{\shortparallel}(t)}
\bR_{\perp}^{\star}(t) -
\norm{\bR_{\shortparallel}^{\star}(t)}\bR_{\mathrm{O}}^{\perp}(t)} \le
2\pi^2\,\abs{\bR_{\mathrm{O}}(t)\cdot\bR^{\star}(t)} \lessapprox 
\Delta\Pi_{\mathrm c} = 2\pi^2 a_{\mathrm{O}}a_{\star}/(1 - e_{\star}) ,
\\[0.7cm]
\displaystyle
\pi^2\norm{\norm{\bR_{\star}^{\shortparallel}(t)}               
\bR^{\perp}_{V}(t) +
\norm{\bR^{\shortparallel}_{V}(t)}\bR^{\star}_{\perp}(t)} \le
2\pi^2\abs{\bR^{\star}(t)\cdot\bR_{V}(t)} \lessapprox \Delta\Psi_c = 
\pi^2V \Delta T a_{\star}/(1 - e_{\star}), \\[0.7cm]
\displaystyle
\pi^2\norm{\norm{\bR_{\mathrm{O}}^{\shortparallel}(t)}               
\bR^{\perp}_{V}(t) -
\norm{\bR^{\shortparallel}_{V}(t)}\bR_{\mathrm{O}}^{\perp}(t)} \le
2\pi^2\,\abs{\bR_{\mathrm{O}}(t)\cdot\bR_{V}(t)} \lessapprox \Delta\Pi_{\mu} 
= \pi^2 a_{\mathrm{O}}V\Delta T
\end{array}
\end{equation}
where $V = (V_T^2 + V_R^2)^{1/2}$. The value of $\Delta\Pi_{\mu}$ has been 
calculated for the same sample of stars as in Fig. 3. Again, one can observe 
that $\Delta\Pi_{\mu}$ is significant and a proper model has to take it into 
account (see Fig.~4). We also find that
\begin{equation}  
  \label{eq:corr6}
\begin{array}{c}
\displaystyle
\Delta \Pi_{\mathrm{c}} \approx 9\times10^{-3}\,\frac{m_{M_{JUP}}
P_{yr}^{2/3}}{d_{pc}^{2}M_{M_{\odot}}^{2/3}(1 - e)}\,\mu as\quad
\mbox{for planetary companions} \\[0.7cm]
\displaystyle
\Delta \Pi_{\mathrm{c}} \approx 9.7 \frac{m_{M_{\odot}}
P_{yr}^{2/3}}{d_{pc}^{2} M_{M_{\odot}}^{2/3} (1 + m_{M_{\odot}}
/M_{M_{\odot}})^{2/3}(1 - e)}\,\mu as
\quad \mbox{for stellar companions}  
\end{array}
\end{equation}
and
\begin{equation}  
  \label{eq:corr7}
\begin{array}{c}
\displaystyle
\Delta \Psi_{\mathrm{c}} \approx 0.1\,V_{100}\Delta T_{yr}\,
\frac{m_{M_{JUP}}
P_{yr}^{2/3}}{d_{pc}^{2}M_{M_{\odot}}^{2/3}(1 - e)}\,\mu as\quad
\mbox{for planetary companions} \\[0.7cm]
\displaystyle
\Delta \Psi_{\mathrm{c}} \approx 102.3\,V_{100}\Delta T_{yr}\, 
\frac{m_{M_{\odot}}
P_{yr}^{2/3}}{d_{pc}^{2} M_{M_{\odot}}^{2/3} (1 + m_{M_{\odot}}
/M_{M_{\odot}})^{2/3}(1 - e)}\,\mu as
\quad \mbox{for stellar companions}  
\end{array}
\end{equation}
where $V_{100}$ is the velocity of star in hundreds of km/s
and $\Delta T_{yr}$ is the time span of the mission in years.

Finally, let us shortly discuss the magnitude of third order terms.
Obviously, they will be detectable only for $\delta\bR$ due to the
proper motion. Thus we can estimate that the resulting angular 
displacement $\delta S^{(3)}$ is
\begin{equation}
\delta S^{(3)} \sim \left(\frac{\norm{\delta\bR}}{\norm{\bR_0}}\right)^3
= \frac{\pi^3}{8}V^3\,\Delta T^3 = 
\frac{\pi^3}{8}(V_R^2 + V_T^2)^{3/2}\,\Delta T^3
\end{equation}
Its value for the sample of stars from Figs.~3-4 is presented in Fig.~5.
As one can see in few cases this third order term can be larger
than $1 \mu as$.

The above analysis clearly shows that a variety of second order effects
and possibly in few cases third order effects will be detectable with SIM. 
Although throughout the rest of this paper we use only the first order 
model presented in sections 2.1-2 to simplify our considerations, in real 
applications a correct model must include higher order 
effects. They can be easily derived given the theoretical background 
presented in section 2.

\section{Orbital motion}

Let us assume that the motion of $N$ planets and their star is described in 
the barycentric system. From the definition of such system, we have the 
following relation for the radius vector of the parent star
\begin{equation}
\label{eq:Rstar}
  \bR^{\star} = -\frac{1}{M_{\star}}\sum_{j=1}^N  
  m_j\bR_j,
\end{equation}
where $\bR_j$ are radius vectors of planets and $M_{\star}$, $m_j$ are the 
mass of the star and the $j$-th planet, respectively. In the first 
approximation, the motion of planets can be described by means of the 
following equations
\begin{equation}
\label{eq:kepler}
  \frac{{\rmd}^2 \bR_j}{{\rmd}t^2} = 
  -{\mu}_j\frac{\bR_j}{\lvert|\bR_j\rvert|^3}, 
       \qquad j = 1,\ldots ,N,  
\end{equation}
where
\[
\mu_j = \frac{GM_{\star}}{(1 + m_j/ M_{\star})^2} .
\]
and the motion of the star can be obtained from the equation \eqref{eq:Rstar}.

\subsection{Elliptic motion and its expansion}
Solutions $\bR_j(t)$ of the equations \eqref{eq:kepler} belong to the family 
of Keplerian orbits among which elliptic orbits are of particular interest 
for farther analysis. Therefore, let us remind their basic properties.

The radius vector  $\bR_j=\bR(t)$ of a planet moving in an elliptic orbit 
is given by
\begin{equation}
\label{e:RtKepler}
   \bR(t) =  \bP\,a(\cos E(t) - e) + \bQ\,a\sqrt{1 - e^2}\sin E(t), 
\end{equation}
where 
\[
\bP = \bl\,\cos\omega + \bm\,\sin\omega,  
\qquad
  \bQ = -\bl\,\sin\omega + {\bf  
m}\cos\omega,
\]
\[
  \bl = \begin{bmatrix}\cos\Omega \\
                     \sin\Omega \\
                       0
\end{bmatrix}, \qquad
  \bm = \begin{bmatrix}-\cos i\sin\Omega \\
                     \phantom{-} \cos i\cos\Omega \\
                     \phantom{- \cos \Omega} \sin i 
         \end{bmatrix}.
\]
The eccentric anomaly $E=E(t)$ is an implicit function of time 
through the Kepler equation
\begin{equation}
\label{e:Kepler}
  E - e\sin E = \cM,
\end{equation}
where $\cM$ is the mean anomaly
\begin{equation}
\label{e:mean}
\cM = n(t - T_{\mathrm{p}}),\qquad n = \frac{2\pi}{P},
\end{equation}
and $P$ is the orbital period of a planet. The remaining parameters
$a,e,\omega,\Omega,T_p$ are the standard Keplerian elements ---
semi-major axis, eccentricity, longitude of pericenter, longitude
of ascending node and time of pericenter.

The functions $\cos E$ and $\sin E$ are periodic with respect to $\cM$ 
and can be expanded in the Fourier series
\begin{equation}
\label{e:cosE}
\begin{split}
  \cos E &= -\frac{1}{2}e +
  \sum_{ k \in {\cal Z}_0 } \frac{1}{k}J_{k-1}(ke)\cos (k\cM), \\
  \sin E &= \sum_{ k \in {\cal Z}_0 } \frac{1}{k}J_{k-1}(ke)\sin
  (k\cM),
\end{split}  
\end{equation}
where $J_n(z)$ is a Bessel function of the first kind of order $n$ and
argument $z$; ${\cal Z}_0$ denotes the set of all positive and
negative integers excluding zero, and $e \in [0,1)$. Thus, using the
equations \eqref{e:RtKepler} and \eqref{e:cosE}, we obtain
\begin{equation}
\label{e:RtFourier}
  \widehat\bR(t) =  {\widehat\bR}^0  + 
   \bA\sum_{ k \in {\cal Z}_0 } \frac{1}{k}J_{k-1}(ke)\cos  
(k\cM) +  
\bB\sum_{ k \in {\cal Z}_0 }\frac{1}{k}J_{k-1}(ke)\sin  
(k\cM),
\end{equation}
where
\[
\widehat\bR(t) = \frac{1}{a}\bR, \quad \widehat \bR^0 =
-\frac{3}{2}\bP\,e, \qquad \bA = \bP \qquad \bB = \bQ \sqrt{1 - e^2}.
\]
It can be written in the following complex form 
\begin{equation}
\label{eq:Rtcomplex}
 \widehat\bR(t) =  \widehat\bR^0   + 
  \sum_{ k \in {\cal Z}_0 }\bTheta_k \,\rme^{\rmi  k\cM}, 
\end{equation}
where
\begin{equation}
\label{eq:Theta}
\bTheta_k = \frac{1}{2k}\left( F_-(k,e)   \bA 
 - \rmi F_+(k,e)  \bB
  \right),
\end{equation}
and
\begin{equation}
  \label{eq:fpm}
  F_\pm(k,e) = J_{k - 1}(ke) \pm J_{k + 1}(ke) . 
\end{equation}
Eventually, using \eqref{e:mean} and \eqref{eq:Rtcomplex}, we obtain the
Fourier expansion of $\widehat\bR(t)$
\begin{equation}
\label{eq:Rtfin}
  \widehat\bR(t) = \widehat\bR^0  + \sum_{ k \in {\cal Z}_0 }{\bLambda}_k \,
\rme^{\rmi knt}, \quad\text{where}\quad
 \bLambda_k = \bTheta_k \rme^{-{\rmi }k n T_{\mathrm{p}}}.
\end{equation}

Let us define the following quantity 
\begin{equation}
\label{e:amp}
 \cA_{k}^{l} = \frac{ \left| \Lambda_{k+1}^{l} \right| }%
                 { \left| \Lambda_{k}^{l}   \right| }, \quad\text{for}
\quad l = 1,2,3, \quad\text{and}\quad k>0,
\end{equation}
i.e. the ratio of amplitudes of two successive harmonics, where $\Lambda^i_j$ 
is the $i$-th component of vector $\bLambda_j$. From the properties of Bessel 
functions we have
\begin{equation}
\label{e:amp2}
  \cA_{k}^{l}(e) = \frac{k}{k+1}  
               \sqrt{\frac{ e^2({A^{l}})^2[J_{k+1}'((k+1)e)]^2
                   + ({B^{l}})^2\left[J_{k+1}((k+1)e)\right]^2 }
                    { e^2({A^{l}})^2\left[J_{k}'(ke)\right]^2
                   + ({B^{l}})^2\left[J_{k}(ke)\right]^2 }},
\end{equation} 
where $J_{n}'(z)$ indicates the derivative of a Bessel function
$J_{n}(z)$ with respect to $z$. It can be proved that for all 
$e \in (0,1)$, $l \in\{1,2,3\}$ and $k>0$ we have $\cA_{k}^{l}(e)<1$.
It means that the expansion of $\widehat\bR(t)$ has an important 
property---moduli of successive harmonics of each of coordinates of 
$\widehat\bR(t)$ decrease strictly monotonically with $k$. 

\subsection{Real expansion}

Given the equations from the previous section, it is possible to
derive the real expansion for every component of the vector
$\widehat\bR(t) =
(\widehat{R}_1(t),\widehat{R}_2(t),\widehat{R}_3(t))$. Namely, we
can express this vector in the form
\begin{equation}
\label{eq:real}
\widehat\bR(t) = \widehat\bR^0
+ \sum_{k = 1}^{\infty}\left(\bC^{k}\cos(knt) + 
\bS^{k}\sin(knt)\right)
\end{equation}
which is more convenient in numerical applications. Using
\eqref{eq:Rtcomplex}, \eqref{eq:Theta} and \eqref{eq:Rtfin} we find
\begin{equation}
\label{eq:CS}
\begin{split}
  \bC^{k} & = \frac{1}{k}\left[ \bP F_-(k,e)\cos( knT_{\mathrm{p}}) - \bQ
    \sqrt{1-e^2}F_+(k,e)
    \sin( knT_{\mathrm{p}})\right], \\
  \bS^{k} & = \frac{1}{k}\left[ \bP F_-(k,e)\sin( knT_{\mathrm{p}}) + \bQ
    \sqrt{1-e^2} F_+(k,e) \cos( knT_{\mathrm{p}})\right].
\end{split}
\end{equation}
From the above formulae immediately follows that amplitudes of
successive harmonics are given by
\begin{equation}
  \label{eq:aml}
 (D_l^k)^2 = (C_l^k)^2 + (S_l^k)^2 = \frac{1}{k^2}
\left[ P_l^2 F_-(k,e)^2 + Q_l^2(1-e^2)F_+(k,e)^2\right], \qquad l = 1,2,3.
\end{equation}
In applications it is convenient to have these expressions in an explicit 
form
\begin{equation}
\begin{split}
  (D_1^k)^2 &= \frac{1}{k^2}\left[F_-(k,e)^2 \left(1
      -\sin^2i\sin^2\Omega\right) + \ F(k,e)
    \left(\cos\Omega\sin\omega + \cos i\sin\Omega\cos\omega\right)^2\right]\\
  (D_2^k)^2 &= \frac{1}{k^2}\left[F_-(k,e)^2 \left(1
      -\sin^2i\cos^2\Omega\right) + \ F(k,e)
    \left(\sin\Omega\sin\omega - \cos i\cos\Omega\cos\omega\right)^2\right]\\
  (D_3^k)^2 &= \frac{1}{k^2}\left[F_-(k,e)^2 + \ F(k,e) \cos^2\omega
  \right]\sin^2i,
\end{split}
\end{equation}
where
\[
F(k,e)= (1-e^2) F_+(k,e)^2 -F_-(k,e)^2.
\]

\subsection{Approximate formulae for small and moderate eccentricities}

Since small and moderate eccentricities are more probable it is useful
to have approximations of the expressions from the previous section.
Namely, using known expansions for Bessel functions we obtain the 
following formulae
\begin{equation}
  \label{eq:powe}
   F_\pm(k,e) = \frac{1}{(k-1)!}\left(\frac{ke}{2}\right)^{k-1} + 
    {\cal O}(e^{k+1}), 
  \quad
 \sqrt{1-e^2}F_+(k,e) =  \frac{1}{(k-1)!}\left(\frac{ke}{2}\right)^{k-1}
   + {\cal O}(e^{k+1})
\end{equation}
Subsequently
\begin{equation}
\label{eq:CS1}
\begin{split}
  \bC^{k} & = \frac{1}{k(k-1)!} \left(\frac{ke}{2}\right)^{k-1} \left[ \bl
    \cos \tilde \omega_k + \bm \sin \tilde\omega_k \right]
  + {\cal O}(e^{k+1}), \\
  \bS^{k} & = \frac{1}{k(k-1)!} \left(\frac{ke}{2}\right)^{k-1} \left[-
    \bl \sin \tilde \omega_k + \bm \cos \tilde\omega_k \right]
  + {\cal O}(e^{k+1}).
\end{split}
\end{equation}
where $\tilde\omega_k = \omega - knT_p$. Finally, we obtain the expansions 
for the amplitudes
\begin{equation}
\begin{split}
  (D_1^k)^2 &=
\left[ \frac{1}{k(k-1)!} \left(\frac{ke}{2}\right)^{k-1}\right]^2 
 \left(1 -\sin^2i\sin^2\Omega\right)  + {\cal O}(e^{2k}), \\
  (D_2^k)^2 &= 
\left[ \frac{1}{k(k-1)!} \left(\frac{ke}{2}\right)^{k-1}\right]^2 
 \left(1 -\sin^2i\cos^2\Omega\right)  + {\cal O}(e^{2k}), \\
  (D_3^k)^2 &=
\left[ \frac{1}{k(k-1)!} \left(\frac{ke}{2}\right)^{k-1}\right]^2 \sin^2i  +
                   {\cal O}(e^{2k}).
\end{split}
\end{equation}
From the above we can obtain the harmonic expansion for a circular
orbit. Namely we find that $\widehat\bR^0=\boldsymbol{0}$ and 
$\bC^{k}=\bS^{k}=\boldsymbol{0}$ for $k>1$. While for $k=1$
\begin{equation}
\label{eq:CSk1}
\begin{split}
  \bC^{k} & =  \left[ \bl
    \cos \tilde \omega + \bm \sin \tilde\omega \right], \\
  \bS^{k} & =  \left[-
    \bl \sin \tilde \omega + \bm \cos \tilde\omega \right] .
\end{split}
\end{equation}
where $\tilde\omega = -nT_{\mathrm{p}}$.

These equations will be especially useful for deriving
orbital elements from the coefficients $\bC^{k}, \bS^{k}$ obtained 
through the analysis of observations. We discuss this issue in the 
next section.

\section{Data analysis}

For the tests we assume the following SIM observing scenario.
At the moments $t_i$ and $t_{i+1}$ the relative delay between the target 
and reference star is measured for two orthogonal baseline orientations 
$\bB_1$ and $\bB_2$. Such measurement gives a two dimensional delay vector 
$\bD_i = (D_1(t_i),D_2(t_{i+1}))$ and is repeated $N$ times over the time span 
of the mission, $\Delta T$. As a result we obtain a two dimensional time 
series ${\cal D} = \{\bD_i, i=1,\ldots,N\}$. The goal of the  
data analysis is to detect planetary signatures in ${\cal D}$ and derive 
the orbital parameters of planets. We solve this problem in two steps.
First we perform Frequency Decomposition (FD) of the time series ${\cal D}$.
The aim of this step 
is to understand the basic properties of ${\cal D}$
i.e. determine the number of planets and estimate their orbital parameters.
The second step is the least-squares analysis based on a specific physical
model established in the previous step. Its is aim is to obtain accurate 
values of the orbital elements and their uncertainties. These two steps are 
described in the following sections. 

\subsection{Harmonic model}

From the theoretical considerations of sections 2 and 3 it follows that
relative delays can be modeled by means of the following expression  
\begin{equation}
\label{eq:fourier}
\begin{array}{c}
\displaystyle
  \bD = {\widehat \bD}^0 + {\widehat \bD}^\mu\,t + \sum_{k=1}^\infty\left[ 
    {\widehat \bC}^{\pi,k}\cos(n_\mathrm{O}k t)
    +  {\widehat \bS}^{\pi,k} \cos(n_\mathrm{O}k t)\right] +  \\[0.7cm]
\displaystyle
  + \sum_{j=1}^N \sum_{k=1}^\infty\left[ {\widehat \bC}^{j,k}\cos(n_j k t) +
    {\widehat \bS}^{j,k} \cos(n_j k t)\right],
\end{array}
\end{equation}
where $N$ denotes the number of planets, $n_{\mathrm{O}}$ and $n_j$
denote the mean motion of SIM and $j$-th planet, respectively. Such 
equation comes directly from the fact that the motion of the 
interferometer and the motion of planets can be expanded into the
Fourier series. Consequently, the above formula is used to describe 
${\cal D}$ and special numerical algorithm is used to obtain the parameters 
\begin{equation}
\label{eq:hpar}
\begin{array}{c}
\displaystyle
     {\widehat \bD}^0 = \begin{bmatrix}
         {\widehat D}^0_1\\ {\widehat D}^0_2
         \end{bmatrix}, \quad
    {\widehat \bD}^\mu =\begin{bmatrix}
         {\widehat D}^\mu_1\\ {\widehat D}^\mu_2
         \end{bmatrix}, \quad  
    {\widehat \bC}^{\pi,k} = \begin{bmatrix}
         {\widehat C}^{\pi,k}_1\\ {\widehat C}^{\pi,k}_2
         \end{bmatrix}, \quad  
 {\widehat \bS}^{\pi,k} = \begin{bmatrix}
         {\widehat S}^{\pi,k}_1\\ {\widehat S}^{\pi,k}_2
         \end{bmatrix}, \\[1cm]
\displaystyle
   {\widehat \bC}^{j,k} = \begin{bmatrix}
         {\widehat C}^{j,k}_1\\ {\widehat C}^{j,k}_2
         \end{bmatrix}, \qquad  
   {\widehat \bS}^{j,k} =\begin{bmatrix}
         {\widehat S}^{j,k}_1\\ {\widehat S}^{j,k}_2
         \end{bmatrix}, \quad n_j, \quad n_O, \quad\quad j=1,\ldots,N, \quad
 k= 1,\ldots, K_j,
\end{array}
\end{equation}
This algorithm has been described in great detail in \citet{Konacki:99a::}.
Let us only note here that in practice, due to limited accuracy of
measurements and the fact that the amplitudes of subsequent harmonics
decrease monotonically, the expansions of \eqref{eq:fourier} are finite and 
the (finite) number of harmonics $K_j$ depends mainly on orbital 
eccentricities and measurement errors. Using our algorithm we can determine 
the number of planets $N$, the number of detectable harmonics $K_j$ and 
determine the basic frequencies and coefficients of \eqref{eq:fourier}. 

In fact, we can assume that the mean motion of SIM $n_\mathrm{O}$ as well
as the other elements of its orbit are known.
In other words $\bR_O(t)$ is known and we can use the following 
more constrained version of the formula \eqref{eq:fourier}
\begin{align}
  \label{eq:fourier1}
  \bD = {\widehat \bD}^0 &+ {\widehat \bD}^\mu\,t + \widehat{\mathbb{D}}^\pi
    \cdot\bR_{\mathrm{O}}(t) +
   \sum_{j=1}^N \sum_{k=1}^{K_j}\left[ {\widehat \bC}^{j,k}\cos(n_j k t) +
    {\widehat \bS}^{j,k} \cos(n_j k t)\right],
\end{align}
This way instead of several parameters ${\widehat \bC}^{\pi,k}, 
{\widehat \bS}^{\pi,k}, n_O$ we have six parameters since
\begin{equation}
\widehat{\mathbb{D}}^\pi = \begin{bmatrix}
         {\widehat D}^\pi_{11}, {\widehat D}^\pi_{12}, {\widehat D}^\pi_{13}\\
         {\widehat D}^\pi_{21}, {\widehat D}^\pi_{22}, {\widehat D}^\pi_{23}
         \end{bmatrix}
\end{equation}

In order to have a better understanding of the parameters of
\eqref{eq:fourier1} let us express them by means of the quantities introduced 
in section 2. Using \eqref{eq:reld}, \eqref{eq:D} and \eqref{eq:Rtfin} we 
find that
\begin{equation}
\label{eq:D0}
\begin{array}{c}
\displaystyle
{\widehat \bD}^0 = 
\begin{bmatrix}
         {\widehat D}^0_1\\ {\widehat D}^0_2
         \end{bmatrix} =
\begin{bmatrix}
         D^0_1\\ D^0_2
         \end{bmatrix} 
- \pi_1 \sum_{j=1}^N  \frac{m_j}{M_\star}a_j
          \begin{bmatrix}
            B_1\, \widehat{R}^{0,j}_1\\ B_2\, \widehat{R}^{0,j}_2
         \end{bmatrix},  
 \\[1cm]
\displaystyle
\widehat{\bD}^\mu =  \begin{bmatrix}
         \widehat{D}^\mu_1\\ \widehat{D}^\mu_2
         \end{bmatrix} =
\begin{bmatrix}
         D^\mu_1\\ D^\mu_2
         \end{bmatrix}, \qquad
\widehat{\mathbb{D}}^\pi = \begin{bmatrix}
         {\widehat D}^\pi_{11}, {\widehat D}^\pi_{12}, {\widehat D}^\pi_{13}\\
         {\widehat D}^\pi_{21}, {\widehat D}^\pi_{22}, {\widehat D}^\pi_{23}
         \end{bmatrix} =
\begin{bmatrix}
\bD^{\pi}_1 \\
\bD^{\pi}_2
\end{bmatrix}
\end{array} 
\end{equation}
where $D^0_i$, $D^\mu_i$ and $ \bD^\pi_i$ are the quantities defined by 
\eqref{eq:D} and calculated for $\bB=\bB_i$, $i=1,2$; ${\widehat{\bR}}^{0,j}$ 
denotes ${\widehat{\bR}}^0$ in the expansion \eqref{eq:real} for the orbit 
of the $j$-th planet and $a_j$ is the semi-major axis of the $j$-th planet. 
The coordinates of ${\widehat{\bR}}^{j} = (\widehat{R}^{j}_1,
\widehat{R}^{j}_2,\widehat{R}^{j}_3)$ are 
expressed in the local frame $\{\be_\alpha,\be_\delta,\be_r\}$. This way 
\begin{equation}
\begin{array}{c}
\displaystyle
\bd^c\cdot\widehat{\bR}^{j} = \pi_1(\bB_1\cdot\be_\alpha)\widehat{R}^{j}_1
 = \pi_1 B_1\widehat{R}^{j}_1, \quad \mbox{for $\bB_1$}\\[0.3cm]
\displaystyle
\bd^c\cdot\bR^{j} = \pi_1(\bB_2\cdot\be_\delta)\widehat{R}^{j}_2 = 
\pi_1 B_2\widehat{R}^{j}_2, \quad \mbox{for $\bB_2$}
\end{array}
\end{equation}
since $\bB_1 = B_1\,\be_\alpha$, $\bB_2 = B_2\,\be_\delta$ where $B_i$ is the 
length of the baseline vector $\bB_i$. Similarly
we have
\begin{equation}
  \label{eq:csp}
   \widehat{\bC}^{j,k}= -\pi_1  \frac{m_j}{M_\star}a_j
          \begin{bmatrix}
            B_1\, C^{j,k}_1\\ B_2\, C^{j,k}_2
         \end{bmatrix}, \qquad
 \widehat{\bS}^{j,k}= -\pi_1 \frac{m_j}{M_\star}a_j
          \begin{bmatrix}
            B_1\, S^{j,k}_1\\ B_2\, S^{j,k}_2
         \end{bmatrix},
\end{equation}
where $\bC^{j,k} = (C^{j,k}_1,C^{j,k}_2,C^{j,k}_3)$ and $\bS^{j,k} =
(S^{j,k}_1,S^{j,k}_2,S^{j,k}_3)$ are $\bC^k$ and $\bS^k$ coefficients
of expansion \eqref{eq:real} for $j$-th planet expressed in the local
frame.

Now let us assume that after performing FD, we obtained the parameters
\begin{equation}
\widehat{\bD}^0, \quad \widehat{\bD}^\mu, \quad \widehat{\mathbb{D}}^\pi,
\quad \widehat{\bC}^{j,k}, \quad \widehat{\bS}^{j,k}, \quad 
j=1,\ldots,N, \quad k=1,\ldots,K_j
\end{equation}
where $N$ is the number of planets (i.e. the number of basic frequencies
detected) and $K_j$ is the number of detected harmonics for each planet.
The first question is if we can derive the canonical parameters like
$\alpha,\delta,\mu_\alpha,\mu_\delta,\pi$  for the target and reference
star from $\widehat{\bD}^0, \widehat{\bD}^\mu, \widehat{\mathbb{D}}^\pi$. 
Unfortunately this is not possible, at least without additional assumptions.
Obviously it is a direct consequence of the relative measurements we
perform. Thus we can only derive $(\bS_0^1 - \bS_0^2)$ as well as
differential proper motion and differential parallax. In fact it is
possible to chose such a reference star that the differential
parallactic displacement has an amplitude close to zero. It suffice to 
have a reference star with the parallax similar to the parallax of the 
target star since by assumption these two stars are close to each other 
and their parallactic displacement is very similar. This way we can remove 
a strong parallactic component from our observations. On the other hand
we do not need the exact values of the canonical parameters. We only
have to properly remove the respective effects in order to be able
to detect putative planets.

The remaining question is if we can derive the orbital elements from
$\widehat{\bC}^{j,k},\widehat{\bS}^{j,k}$. This task is relatively simple.
Namely, given that we have detected at least two terms (basic frequency
and its first harmonics), all orbital elements can be derived from
the equations of section 3.3. For planets with only the basic frequency
detectable, we assume a circular orbit and then the other elements can also
be found. This procedure we demonstrate in section 5.

\subsection{Standard model}

The harmonic model allows us to describe the data without a priori knowledge
of the target star parameters and its planetary system. In the same time it 
allows to derive all important information --- especially the number of planets
and estimates of their orbital elements. With such knowledge we are ready
to perform the standard least-squares analysis in which we must specify
the model and supply good initial conditions for the fit. 

The standard model has the following form
\begin{align}
  \label{eq:std}
  \bD = {\widehat \bD}^0 &+ {\widehat \bD}^\mu\,t + \widehat{\mathbb{D}}^\pi
    \cdot\bR_{\mathrm{O}}(t) +
   \sum_{j=1}^N 
          \begin{bmatrix}
            \widehat{a}_j\,B_1\,\widehat{R}_1(t,T_{\mathrm{p},j}, e_j, i_j, \omega_j,
\Omega_j,  P_j) \\ \widehat{a}_j\,B_2\,\widehat{R}_2(t,T_{\mathrm{p},j}, e_j, i_j,
\omega_j, \Omega_j,  P_j) 
         \end{bmatrix},
\end{align}
where $\widehat{R}_1, \widehat{R}_2$ are coordinates of the Keplerian motion 
vector $\widehat{\bR}$ given by the equation \eqref{e:RtFourier}. The 
parameters of such model are
 \begin{equation}
   \label{eq:hpar1}
     \bD^0, \quad\bD^\mu,\quad{\mathbb D}^\pi, \quad
   \widehat{a}_j, T_{\mathrm{p},j}, e_j, i_j, \omega_j, \Omega_j,  P_j,
    \quad j=1,\ldots,N,
 \end{equation}
where $T_{\mathrm{p},j},e_j, i_j, \omega_j, \Omega_j$ are the Keplerian 
elements of the $j$-th planet, $P_j$ is its orbital period and the 
parameter $\widehat{a}_j$ is defined in the following way
\begin{equation}
\label{eq:ta}
\widehat{a}_j = \pi_1 a_j \frac{m_j}{M_\star}
\end{equation}

\section{Numerical tests}

For the tests we chose $\upsilon$ And with its two outer planets 
\citep{Butler:99::}. All real and assumed astrometric and orbital 
parameters are presented in Table 1. We also found a reference
star HD~10032 which is about $0. \!^\circ 7$ away from $\upsilon$ And. 
Its astrometric parameters are in Table 2. SIM is assumed to move in
an orbit similar to the orbit of the Earth (see Table 3). For these
stars we simulated $N = 200$ measurements of relative delays $(D_1,D_2)$ 
(for two baseline vector orientations $\bB_1$ and $\bB_2$) randomly 
distributed over the time span of 10 years. In both cases the length of the 
baseline vector was 10 meters and a measurement error with 
$\sigma \approx 50$ pm was assumed (i.e. $1 \mu as$ in angular 
displacement). Since by assumption $\bB_1$ is parallel to $\be_\alpha$ and 
$\bB_2$ to $\be_\delta$, the delay $D_1$ corresponds to an angular
distance between $\upsilon$ And and HD~10032 in right ascension and 
$D_2$ to an angular distance in declination.

\subsection{Second order effects}

Before we proceed with the analysis it is interesting to discuss 
second order effects present in simulated observations. Since $\upsilon$
And is a nearby star with large proper motion we can expect
significant contribution from this star (our reference star HD~10032
is quite distant and thus all second order effects are mainly due to
$\upsilon$ And). One can analyze these effects by means of the formulae
from section 2.3 or simply apply the standard model \eqref{eq:std} with
the parameters precisely computed from assumed parameters of $\upsilon$ And,
HD~10032 and SIM (Tables 1-3) and examine the resulting residuals.
This procedure gives the residuals presented in Fig. 6. As we can see
the second order effects are dominated by a variation quadratic in time
(Fig.~6 a,b). This effect is due to perspective acceleration
\begin{equation}   
\frac{\pi^2}{4}\norm{\bV_R}\bV_T\,\Delta T^2 
\end{equation}
thus if we assume that the radial velocity, $\bV_R$, of $\upsilon$ And and
HD~10032 is zero it will disappear and reveal another
second order effect of smaller magnitude (see Fig.~6 c,d). This
effect is due to the following term
\begin{equation}
\pi^2\norm{\bR_{\mathrm{O}}^{\shortparallel}(t)}               
\bR^{\perp}_{V}(t) - \pi^2
\norm{\bR^{\shortparallel}_{V}(t)}\bR_{\mathrm{O}}^{\perp}(t) 
\end{equation}
i.e. the mixed term of parallax and motion of the star.

Finally let us note that if we allow the parameters of the model 
\eqref{eq:std} to vary, as usual during the process of least-squares
fit, this first order model will try to minimize the residuals
as presented in Fig.~6 e,f.

\subsection{Frequency Decomposition and standard model}

First step in our analysis of simulated relative delay measurements
is the Frequency Decomposition (FD). Here we model the data with the less
constrained formula \eqref{eq:fourier} to show how the parallactic motion contributes
to the data. From assumed parameters of the stars and SIM we can compute
amplitudes of basic terms and their harmonics. They are shown in Fig. 7.
The main idea of FD is subsequent removal of effects with decreasing
magnitudes (for all details of the method see \citet{Konacki:99a::}). This 
process is demonstrated in Fig. 8 and 9 for $D_1$ (i.e. for delays measured
with the baseline vector orientation $\bB_1$). As one can see the most 
significant part of delay variations comes from the proper motion of both stars (Fig.
8a), then we can detect the basic term of the parallactic motion (Fig. 8b),
the basic term of the planet II (Fig. 8c), first harmonic of the parallactic
motion (Fig. 8d), the basic term of the planet I (Fig. 9e), first harmonic of
the planet II (Fig. 9f), second harmonic of the planet II (Fig. 9g) and
finally first harmonic of the planet I (Fig. 9h). The values of respective
parameters $\widehat{\bS}^{j,k}, \widehat{\bC}^{j,k}$ are presented in
Table 4. They are sufficient to derive initial estimates of the orbital
elements of planets I and II. Namely, from the approximate equations of 
section 3.3 we can find that
\begin{equation}
\label{eq:orb_elem}
\begin{array}{c}
\displaystyle
e_j = 2\,\sqrt{\frac{({\widehat C}^{j,2}_1)^2 + ({\widehat S}^{j,2}_1)^2}
{({\widehat C}^{j,1}_1)^2 + ({\widehat S}^{j,1}_1)^2}}, \\[0.7cm]
\displaystyle
\frac{{\widehat S}^{j,1}_1}{B_1}\,\frac{{\widehat C}^{j,1}_2}{B_2} 
- \frac{{\widehat S}^{j,1}_2}{B_2}\,\frac{{\widehat C}^{j,1}_1}{B_1} 
= -\widehat{a}_j^2\,\cos i_j \\[0.7cm]
\displaystyle
\left(\frac{{\widehat C}^{j,1}_1}{B_1}\right)^2 + 
\left(\frac{{\widehat C}^{j,1}_2}{B_2}\right)^2 -
\left(\frac{{\widehat S}^{j,1}_1}{B_1}\right)^2 -
\left(\frac{{\widehat S}^{j,1}_2}{B_2}\right)^2 = \widehat{a}_j^2\,\cos
2\tilde\omega_{1,j}\,\sin^2 i_j, \\[0.7cm]
\displaystyle
\left(\frac{{\widehat S}^{j,1}_1}{B_1}\right)^2 +
\left(\frac{{\widehat C}^{j,1}_1}{B_1}\right)^2 -
\left(\frac{{\widehat S}^{j,1}_2}{B_2}\right)^2 -
\left(\frac{{\widehat C}^{j,1}_2}{B_2}\right)^2 =\widehat{a}_j^2\,\cos
2\Omega_j\,\sin^2 i_j, \\[0.7cm]
\displaystyle
\frac{{\widehat C}^{j,1}_1}{B_1}\,\frac{{\widehat S}^{j,1}_1}{B_1}
+ \frac{{\widehat C}^{j,1}_2}{B_2}\,\frac{{\widehat S}^{j,1}_2}{B_2} =
-\widehat{a}_j^2\,\sin\tilde\omega_{1,j}\cos\tilde\omega_{1,j}\,
\sin^2 i_j, \\[0.7cm]
\displaystyle
\frac{{\widehat C}^{j,1}_1}{B_1}\,\frac{{\widehat C}^{j,1}_2}{B_2}
+ \frac{{\widehat S}^{j,1}_1}{B_1}\,\frac{{\widehat S}^{j,1}_2}{B_2} =
\widehat{a}_j^2\,\sin\Omega_j\cos\Omega_j\,\sin^2 i_j
\end{array}
\end{equation}
and, together with analogous formulae for first harmonics, easily determine 
the orbital elements. They are show in Table 5. As one can see 
this procedure gives quite accurate values of the orbital
elements. However we use them only as initial values for the least-squares
fit with the standard model \eqref{eq:std} to obtain the final parameters
presented in Table 6.

\subsection{Conclusions}

The above test demonstrates that our approach allows us to determine the
orbital elements with high confidence, at least in this particular case. It
is interesting to note that with FD we are able to estimate the orbital
elements without using the entire information present in the simulated data
set (the residuals from Fig. 9h are well above the assumed measurement
error). Surprisingly this estimation is quite accurate and as demonstrated
is perfectly sufficient as an initial guess of the parameters for the
standard least-squares analysis. This is a very promising result since the
difficult problem of good initial condition is usually solved by means of 
quasi-global techniques which are very demanding numerically and still may 
lead to unreliable results. Thus we believe that our approach constitutes 
safe and efficient solution to the problem of planets detection with SIM. In 
our forthcoming paper we will thoroughly analyze the method on more realistic
simulations and a variety of different planetary systems.

\clearpage

%
%

\figcaption[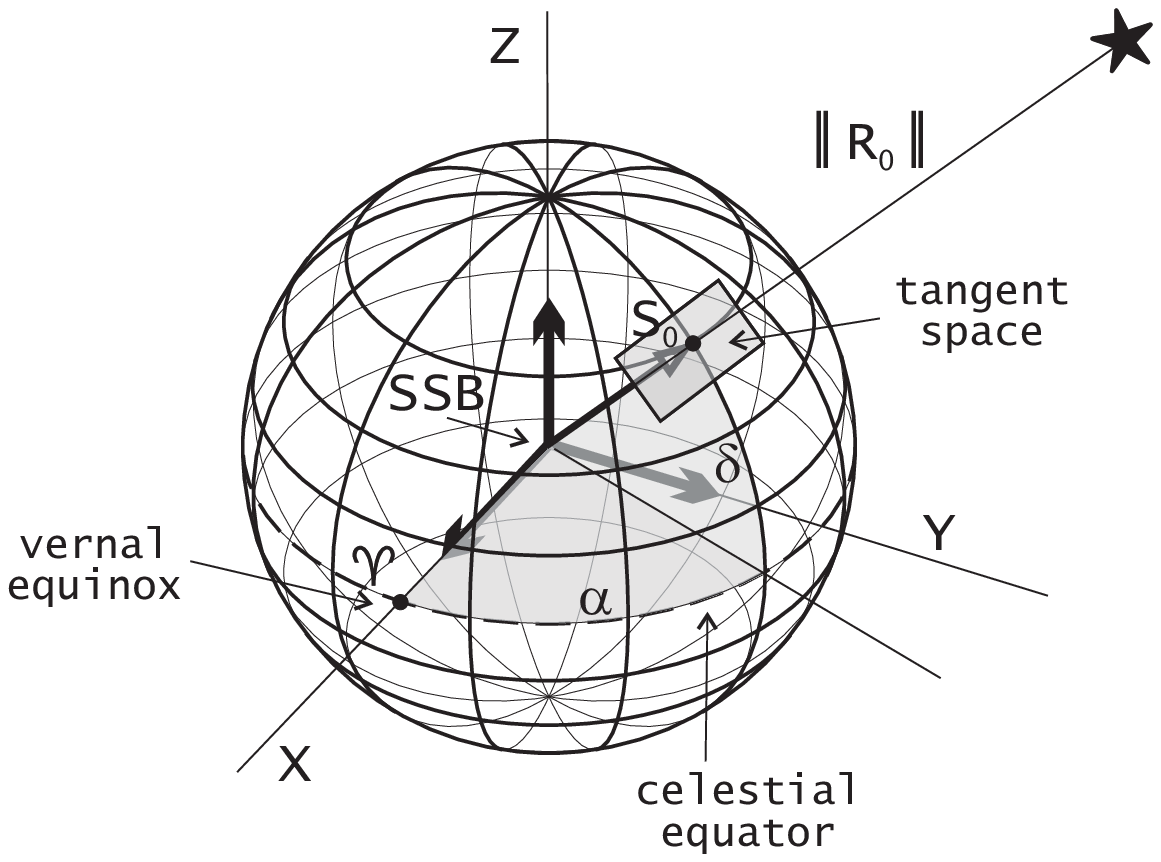]{Solar System Barycenter (SSB) reference frame
and the celestial sphere. $\bS_0$ is the unit vector toward the star
with spherical coordinates $(\alpha,\delta)$ and $\norm{\bR_0}$ is
the distance to the star.}

\figcaption[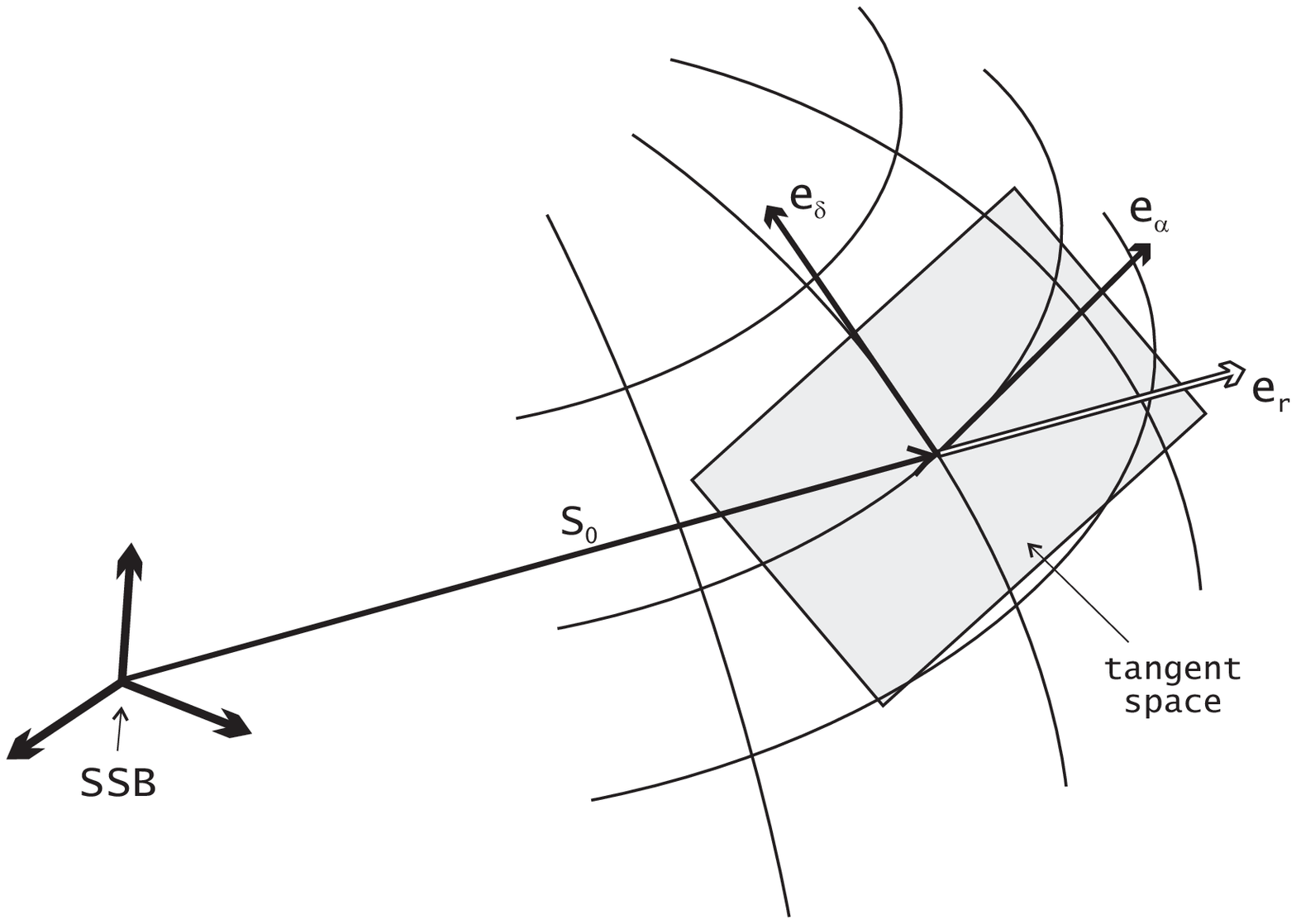]{Tangent space at $\bS_0$ where $\be_{\alpha}$, 
$\be_{\delta}$ and $\be_{r}$ are the unit vectors of the local frame.}

\figcaption[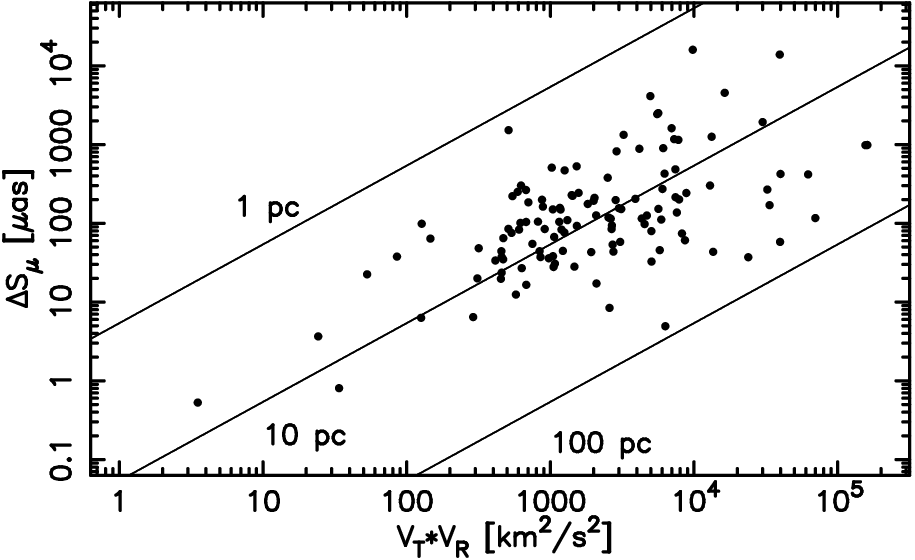]{$\Delta S_{\mu}$ for 150 stars with the largest
proper motion from the Hipparcos catalogue. The solid lines represent 
$\Delta S_{\mu}$ for 1, 10 and 100 parsecs as a function of $V_T\,V_R$. 
Time span of the mission $\Delta T = 10 yr$ was assumed.}

\figcaption[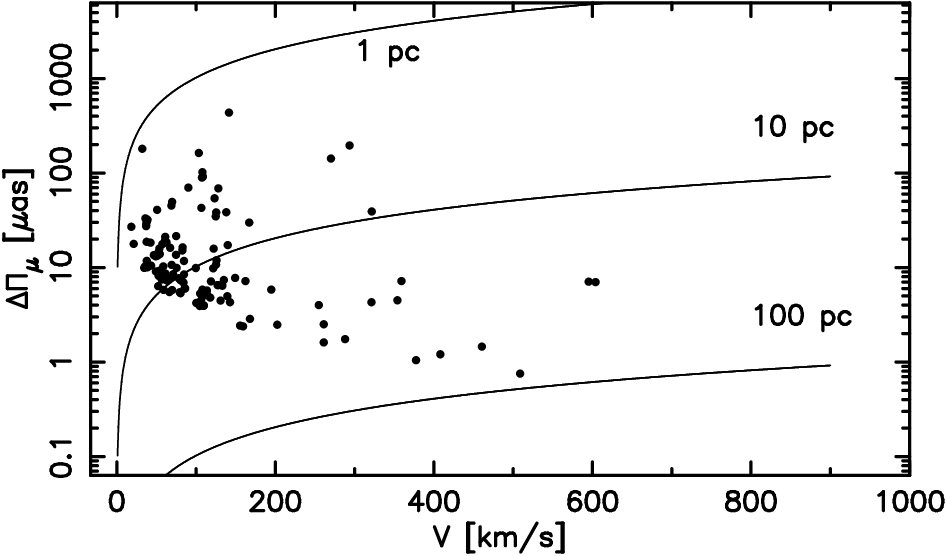]{$\Delta \Pi_{\mu}$ for 150 stars with the largest
proper motion from the Hipparcos catalogue. The solid lines represent 
$\Delta \Pi_{\mu}$ for 1, 10 and 100 parsecs as a function of $V$. 
Time span of the mission $\Delta T = 10 yr$ was assumed.}

\figcaption[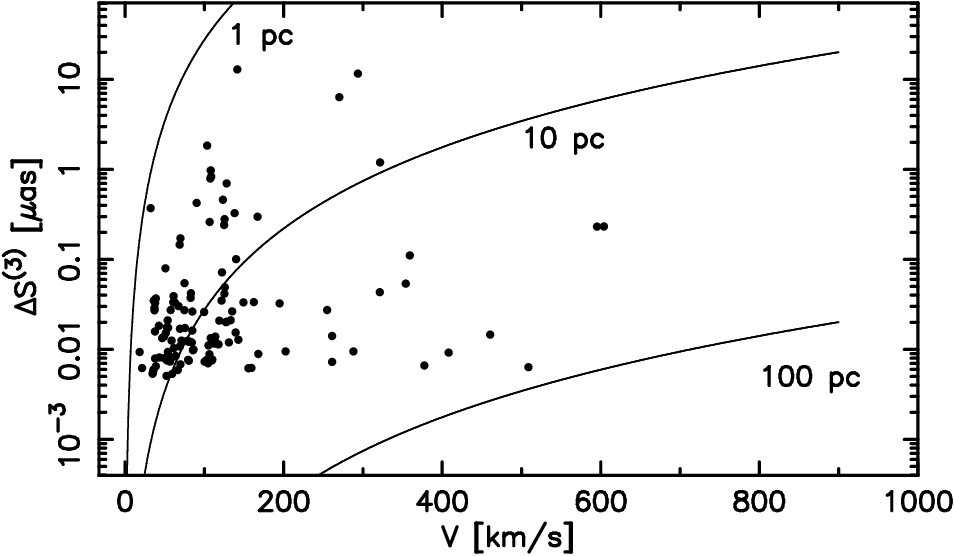]{$\Delta S^{(3)}$ for 150 stars with the largest
proper motion from the Hipparcos catalogue. The solid lines represent 
$\Delta S^{(3)}$ for 1, 10 and 100 parsecs as a function of $V$. Time
span of the mission $\Delta T = 10 yr$ was assumed.}

\figcaption[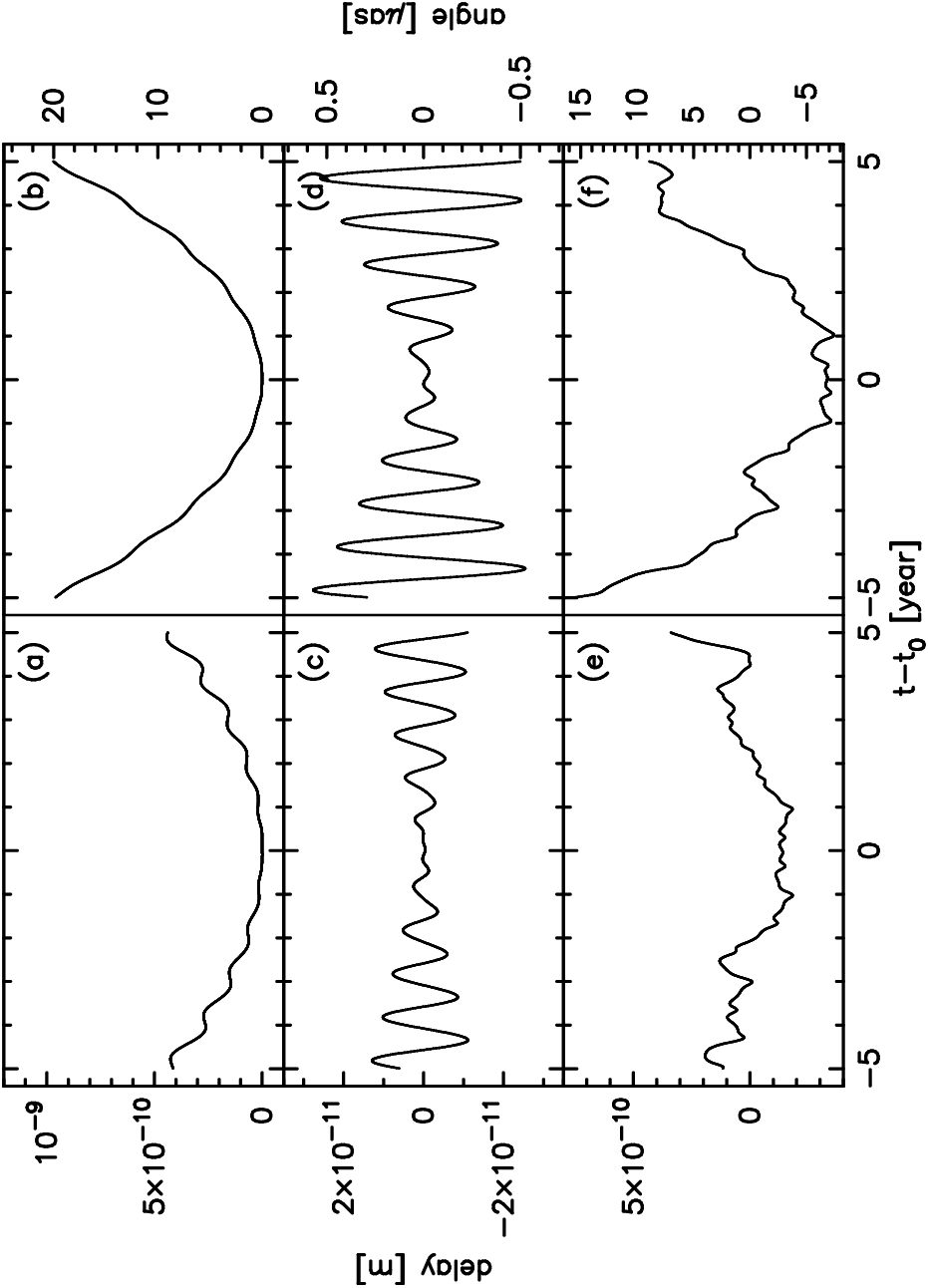]{Second order effects in the simulated delays for 
the relative measurements between $\upsilon$ And and HD~10032 for the
baseline vector orientations $\bB_1$ ({\it a}) and $\bB_2$ ({\it b});
({\it c,d}) the same effects when the radial velocity of both stars
is zero; ({\it e,f}) the residuals from the least-squares fit of the
first order model \eqref{eq:std} to the simulated data used in ({\it a,b}).}

\figcaption[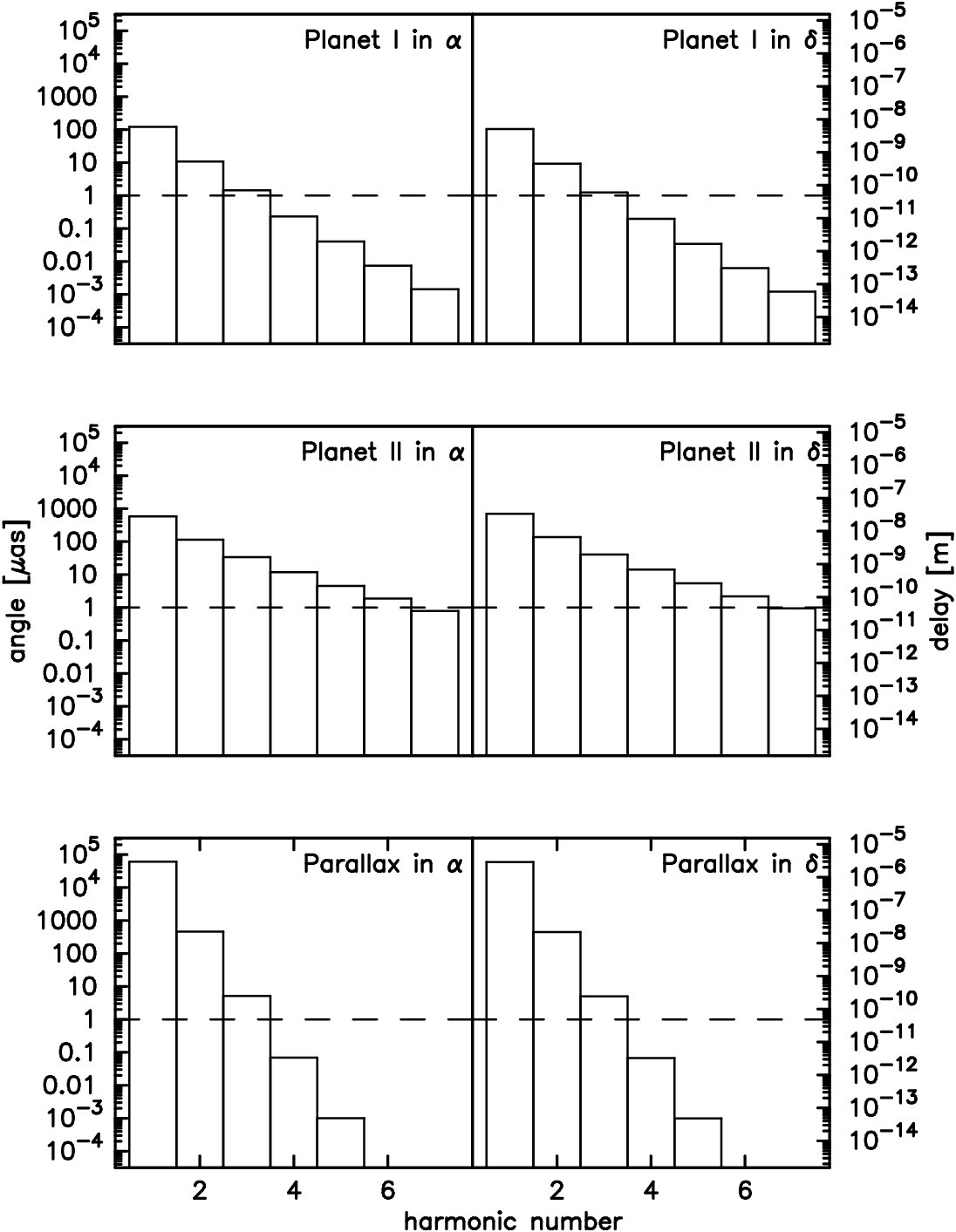]{The amplitudes of subsequent harmonic terms
for the relative delays $D_1,D_2$ (left and right panel
respectively) corresponding to the planet I ({\it a}), II ({\it b})
and the parallactic motion ({\it c}).}

\figcaption[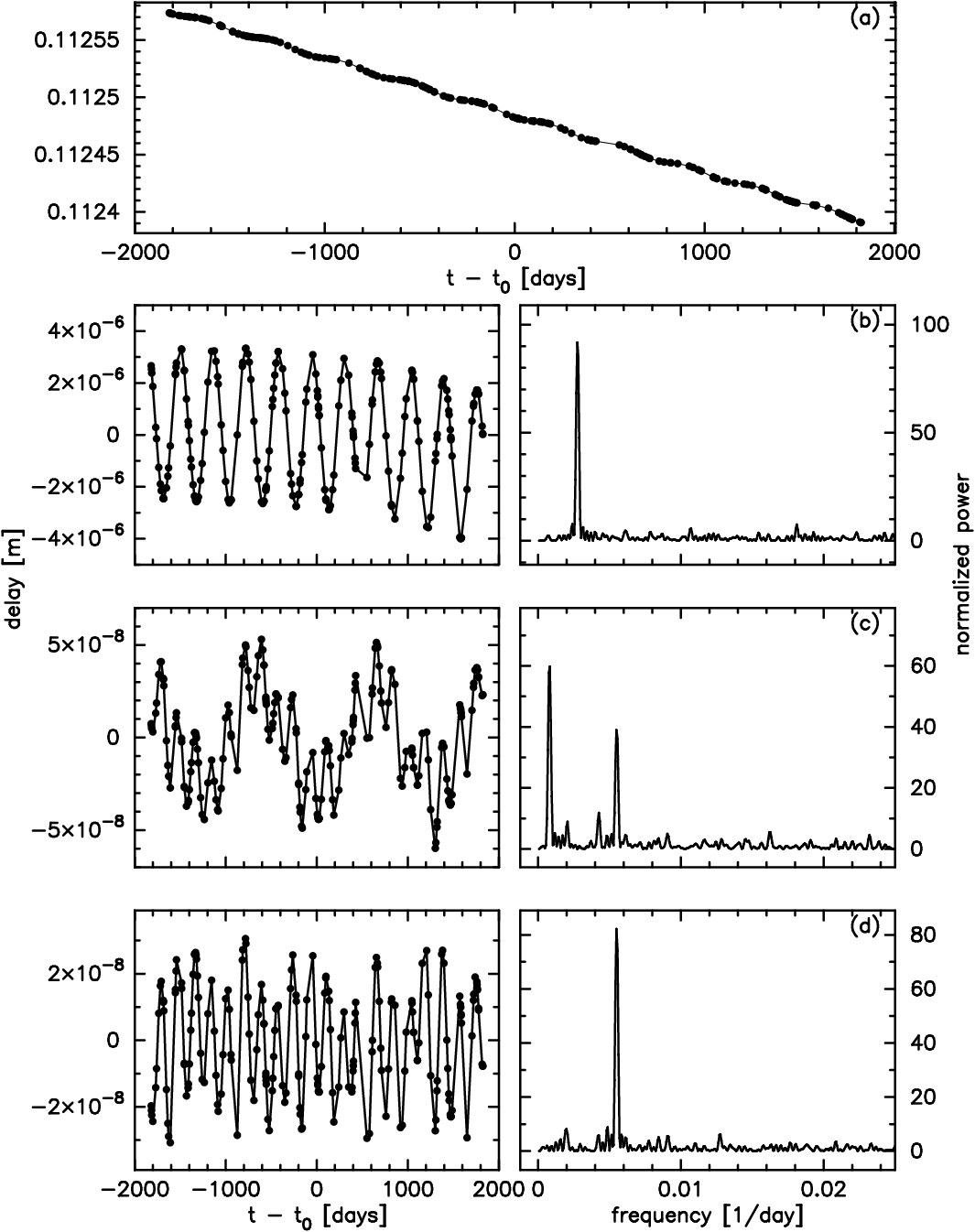]{Subsequent steps of the Frequency Decomposition for
the simulated relative delay measurements between $\upsilon$ And and
HD~10032 corresponding to the baseline vector orientation $\bB_1$.
Left panel contains the residuals after removal of all components
from the steps above. Right panel contains normalized periodograms
of these residuals.}

\figcaption[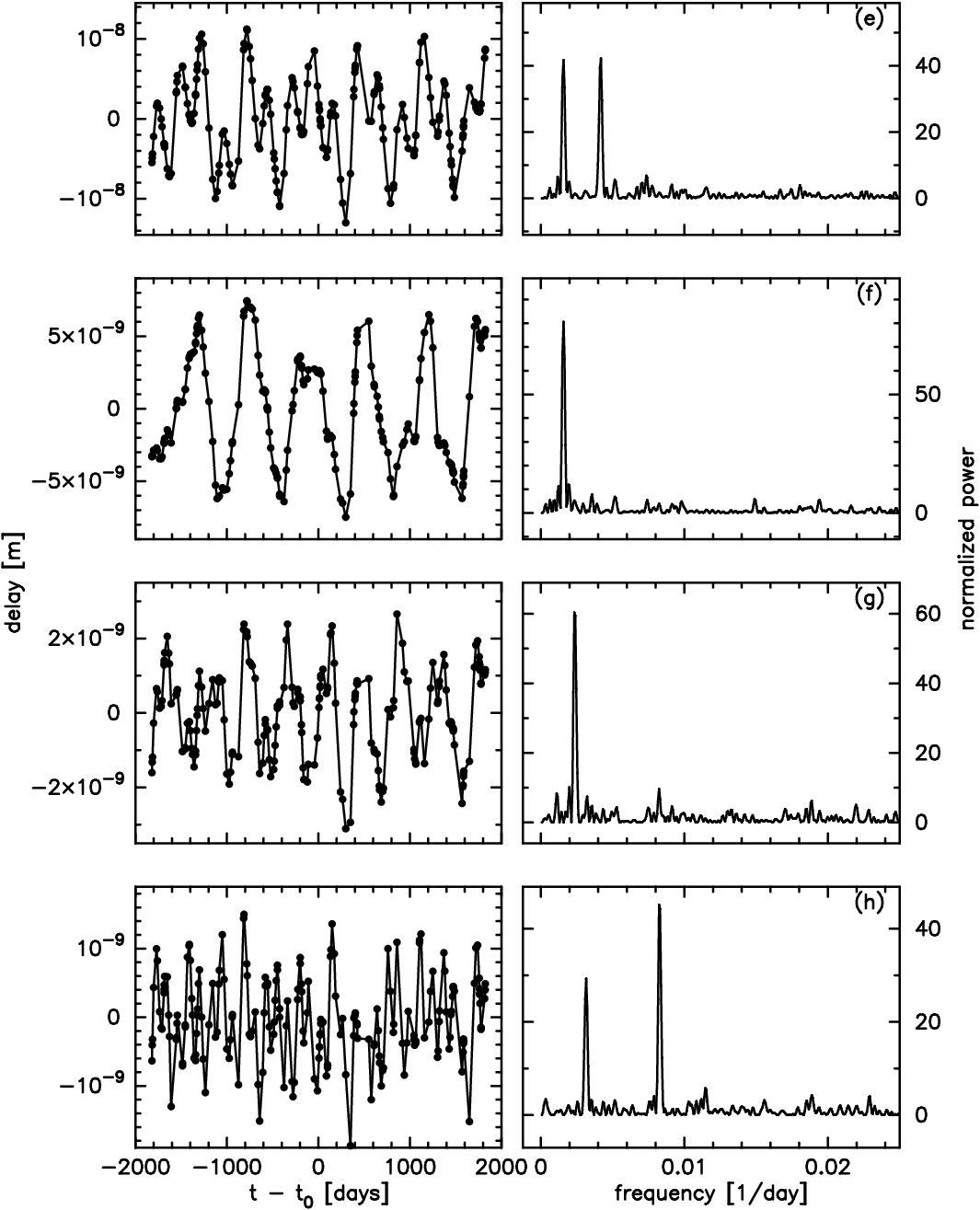]{Continuation of Fig. 8}

\clearpage
 
%
%

%
%

\begin{deluxetable}{lc}
\tabletypesize{\small}
\tablewidth{0pt}
\tablecaption{Target star --- $\upsilon$ Andromedae (HD~9826, HIP~7513)}
\tablehead{
\colhead{Parameter} &
\colhead{$\upsilon$ And} 
}
\startdata
Right ascension, $\alpha$ (J1991.25)\dotfill&
01$^h$36$^m$47.$\!\!^s$98\\              
Declination, $\delta$ (J1991.25)\dotfill& 
41$^\circ$24$^\prime$23.$\!\!^{\prime\prime}$00\\
Proper motion in $\alpha$, $\mu_{\alpha}\cos\delta$ (mas/yr)\dotfill &
-172.57\\
Proper motion in $\delta$, $\mu_{\delta}$ (mas/yr)\dotfill& -381.01\\
Parallax, $\pi$ (mas)\dotfill &  74.25\\
Distance, $d_{pc}$ (pc)\dotfill  & 13.47\\
Transverse velocity, $V_T$ (km/s)\dotfill & 26.7\\
Radial velocity, $V_R$ (km/s)\dotfill & -27.7\\
Mass, $M_{\star}$ ($M_{\odot}$)\dotfill & 1.3 \\
\cutinhead{\phm{xxxxxxxx}Orbital elements\phm{xxxxxxxxxx}Planet I\phm{xxxxx.}Planet II}
Semi-major axis, $a$ (AU)\dotfill &
\multicolumn{1}{l}{0.83 \phm{8888888} 2.5} \\
Semi-major axis, $\widehat{a} = \pi\,a\,m/M_{\star}$ (mas)\dotfill &
\multicolumn{1}{l}{0.133 \phm{888888} 0.813} \\
Orbital period, $P$ (d)\dotfill &
\multicolumn{1}{l}{241.2 \phm{888888} 1266.6} \\
Eccentricity, $e$\dotfill &
\multicolumn{1}{l}{0.18 \phm{8888888} 0.41} \\
Epoch of periastron, $T_p$ (JD)\dotfill &
\multicolumn{1}{l}{2450154.9 \phm{88} 2451308.7} \\
Longitude of periastron, $\omega$\dotfill&
\multicolumn{1}{l}{243.$\!^\circ$6 \phm{88888$\!^\circ$} 247.$\!^\circ$7}\\
Longitude of ascending node, $\Omega$\dotfill&
\multicolumn{1}{l}{30.$\!^\circ$0 \phm{888888$\!^\circ$} 60.$\!^\circ$0}\\
Inclination, $i$\dotfill&
\multicolumn{1}{l}{45.$\!^\circ$0 \phm{888888$\!^\circ$} 45.$\!^\circ$0}\\
Mass, $m$ ($M_{JUP}$)\dotfill & 
\multicolumn{1}{l}{2.95 \phm{8888888} 5.98} \\
\enddata
\end{deluxetable}

%
%

\begin{deluxetable}{lc}
\tablewidth{0pt}
\tablecaption{Reference star --- HD~10032 (HIP~7672)}
\tablehead{
\colhead{Parameter} &
\colhead{HD~10032} 
}
\startdata
Right ascension, $\alpha$ (J1991.25)\dotfill&
01$^h$38$^m$48.$\!\!^s$07\\              
Declination, $\delta$ (J1991.25)\dotfill& 
40$^\circ$45$^\prime$38.$\!\!^{\prime\prime}$80\\
Proper motion in $\alpha$, $\mu_{\alpha}\cos\delta$ (mas/yr)\dotfill &
-14.70\\
Proper motion in $\delta$, $\mu_{\delta}$ (mas/yr)\dotfill& -2.66\\
Parallax, $\pi$ (mas)\dotfill &  8.05\\
Distance, $d_{pc}$ (pc)\dotfill  & 124.22\\
Transverse velocity, $V_T$ (km/s)\dotfill & 8.80\\
Radial velocity, $V_R$ (km/s)\dotfill & -34.00\\
\enddata
\end{deluxetable}

%
%

\begin{deluxetable}{lc}
\tablewidth{0pt}
\tablecaption{SIM orbital elements in SSB reference frame}
\tablehead{
\colhead{Parameter} &
\colhead{SIM} 
}
\startdata
Semi-major axis, $a$ (AU)\dotfill & 0.995 \\
Orbital period, $P$ (d)\dotfill &  362.5\\
Eccentricity, $e$\dotfill & 0.015 \\
Epoch of periastron, $T_p$ (JD)\dotfill & 2451519.44 \\
Longitude of periastron, $\omega$\dotfill& 74.$\!^\circ$67\\
Longitude of ascending node, $\Omega$\dotfill& 0.$\!^\circ$005\\
Inclination, $i$\dotfill& 23.$\!^\circ$45\\
\enddata
\end{deluxetable}

%
%

\begin{deluxetable}{lcc}
\tablewidth{0pt}
\tablecaption{Dominant planetary terms from Frequency Decomposition}
\tablehead{
\colhead{Parameter} &
\colhead{Planet I} &
\colhead{Planet II}
}
\startdata
$f$ (1/d) \dotfill  & 1/241.35  & 1/1265.87 \\
${\widehat C}^{1}_1$ (m) \dotfill  
& $-0.743 \times 10^{-11}$ & $0.282 \times 10^{-7}$ \\
${\widehat S}^{1}_1$ (m) \dotfill  
& $0.586 \times 10^{-8}$ & $0.790 \times 10^{-9}$ \\
${\widehat C}^{1}_2$ (m) \dotfill  
& $-0.481 \times 10^{-8}$ & $0.756 \times 10^{-8}$ \\
${\widehat S}^{1}_2$ (m) \dotfill  
& $0.146 \times 10^{-8}$ & $0.329 \times 10^{-7}$ \\
${\widehat C}^{2}_1$ (m) \dotfill  
& $0.198 \times 10^{-9}$ & $-0.301 \times 10^{-8}$ \\
${\widehat S}^{2}_1$ (m) \dotfill  
& $-0.683 \times 10^{-9}$ & $0.466 \times 10^{-8}$ \\
${\widehat C}^{2}_2$ (m) \dotfill  
& $0.492 \times 10^{-9}$ & $-0.624 \times 10^{-8}$ \\
${\widehat S}^{2}_2$ (m) \dotfill  
& $-0.148 \times 10^{-9}$ & $-0.202 \times 10^{-8}$ \\
\enddata
\end{deluxetable}

%
%

\begin{deluxetable}{lll}
\tablewidth{0pt}
\tablecaption{Orbital elements derived from FD parameters}
\tablehead{
\colhead{Parameter} &
\colhead{Planet I} &
\colhead{Planet II}
}
\startdata
Semi-major axis, $\widehat{a}$ (mas)\dotfill & 0.130 & 0.733 \\
Orbital period, $P$ (d)\dotfill & 241.35 & 1265.87 \\
Eccentricity, $e$\dotfill & 0.22 & 0.39 \\
Epoch of periastron, $T_p$ (JD)\dotfill &  2450164.79 & 2451301.97 \\
Longitude of periastron, $\omega$\dotfill& 255.$\!^\circ$05 &
243.$\!^\circ$15\\
Longitude of ascending node, $\Omega$\dotfill& 31.$\!^\circ$09 &
62.$\!^\circ$95\\
Inclination, $i$\dotfill& 44.$\!^\circ$50 & 43.$\!^\circ$13\\
\enddata
\end{deluxetable}

%
%

\begin{deluxetable}{lll}
\tablewidth{0pt}
\tablecaption{Orbital elements from standard model}
\tablehead{
\colhead{Parameter} &
\colhead{Planet I} &
\colhead{Planet II}
}
\startdata
Semi-major axis, $\widehat{a}$ (AU)\dotfill & 0.132 & 0.813 \\
Orbital period, $P$ (d)\dotfill & 241.21 & 1265.65 \\
Eccentricity, $e$\dotfill & 0.17 & 0.41 \\
Epoch of periastron, $T_p$ (JD)\dotfill & 2450152.63 & 2451306.95 \\
Longitude of periastron, $\omega$\dotfill& 240.$\!^\circ$34 &
247.$\!^\circ$88\\
Longitude of ascending node, $\Omega$\dotfill& 29.$\!^\circ$65 & 
59.$\!^\circ$96\\
Inclination, $i$\dotfill& 44.$\!^\circ$91 & 45.$\!^\circ$04\\
\enddata
\end{deluxetable}

\clearpage

%
%

%
%

\begin{figure}
\figurenum{1}
\epsscale{0.7}
\plotone{f1.eps}
\caption{}
\end{figure}


%
%

\begin{figure}
\figurenum{2}
\epsscale{0.7}
\plotone{f2.eps}
\caption{}
\end{figure}


%
%

\begin{figure}
\figurenum{3}
\epsscale{0.9}
\plotone{f3.ps}
\caption{}
\end{figure}


%
%

\begin{figure}
\figurenum{4}
\epsscale{0.9}
\plotone{f4.ps}
\caption{}
\end{figure}


%
%

\begin{figure}
\figurenum{5}
\epsscale{0.9}
\plotone{f5.ps}
\caption{}
\end{figure}


%
%

\begin{figure}
\figurenum{6}
\epsscale{0.8}
\plotone{f6.ps}
\caption{}
\end{figure}


%
%

\begin{figure}
\figurenum{7}
\epsscale{0.9}
\plotone{f7.ps}
\caption{}
\end{figure}


%
%

\begin{figure}
\figurenum{8}
\epsscale{0.9}
\plotone{f8.ps}
\caption{}
\end{figure}


%
%

\begin{figure}
\figurenum{9}
\epsscale{0.9}
\plotone{f9.ps}
\caption{}
\end{figure}

\end{document}

%% file: mathdef.tex
\newcommand\bb{{\mathbf b}} 
 
\newcommand\bd{{\mathbf d}} 
\newcommand\be{{\mathbf e}}

\newcommand\bl{{\mathbf l}} 
\newcommand\bm{{\mathbf m}}

\newcommand\bs{{\mathbf s}}

\newcommand\bA{{\mathbf A}} 
\newcommand\bB{{\mathbf B}} 
\newcommand\bC{{\mathbf C}} 
\newcommand\bD{{\mathbf D}}

\newcommand\bP{{\mathbf P}}
\newcommand\bQ{{\mathbf Q}} 
\newcommand\bR{{\mathbf R}} 
\newcommand\bS{{\mathbf S}}

\newcommand\bV{{\mathbf V}}

\newcommand\cA{{\mathcal A}}

\newcommand\cM{{\mathcal M}}

\newcommand\bTheta{\boldsymbol{\Theta}}
\newcommand\bLambda{\boldsymbol{\Lambda}}
\newcommand\bPi{\boldsymbol{\Pi}}
\newcommand\bmu{\boldsymbol{\mu}}

\newcommand\rmd{\mathrm{d}}
\newcommand\rmi{\mathrm{i}}
\newcommand\rme{\mathrm{e}}

\newcommand\abs[1]{\lvert #1 \rvert}
\newcommand\norm[1]{\lVert #1 \rVert}